\documentclass[english,aps,prl,twocolumn,amsmath,amssymb,showpacs,superscriptaddress,notitlepage,longbibliography]{revtex4-2}
\usepackage[LGR,T1]{fontenc}
\usepackage[utf8]{inputenc} 
\usepackage{color}
\usepackage{bm}
\usepackage{dsfont}
\usepackage{amsmath}
\usepackage{amssymb}
\usepackage{graphicx}
\usepackage{cancel}

\makeatletter

\DeclareRobustCommand{\greektext}{%
  \fontencoding{LGR}\selectfont\def\encodingdefault{LGR}}
\DeclareRobustCommand{\textgreek}[1]{\leavevmode{\greektext #1}}
\ProvideTextCommand{\~}{LGR}[1]{\char126#1}

\usepackage{amssymb}
\usepackage{amsmath}
\usepackage{graphicx}
\usepackage[colorlinks=true,linkcolor=blue,anchorcolor=red,citecolor=blue,urlcolor=blue]{hyperref}
\usepackage[caption=false]{subfig}
\usepackage{lipsum}

\makeatother

\usepackage{babel}
\begin{document}
\renewcommand{\figurename}{Fig.}
\title{Instanton-Induced Supersymmetry Breaking in Topological Semimetals}
\author{W.\ B. Rui}
\email{wbrui@hku.hk}

\address{Department of Physics and HK Institute of Quantum Science \& Technology,
The University of Hong Kong, Pokfulam Road, Hong Kong, China}

\author{Y.\ X. Zhao}
\email{yuxinphy@hku.hk}
\address{Department of Physics and HK Institute of Quantum Science \& Technology,
The University of Hong Kong, Pokfulam Road, Hong Kong, China}
\address{Hong Kong Branch for Quantum Science Center of Guangdong-Hong Kong-Macau Great Bay Area, Shenzhen, China}

\author{Z.\ D. Wang}
\email{zwang@hku.hk}
\address{Department of Physics and HK Institute of Quantum Science \& Technology,
The University of Hong Kong, Pokfulam Road, Hong Kong, China}
\address{Hong Kong Branch for Quantum Science Center of Guangdong-Hong Kong-Macau Great Bay Area, Shenzhen, China}

\date{\today}
\begin{abstract}
Supersymmetry (SUSY) proposed as an elementary symmetry for physics beyond
the Standard Model has found important applications in various areas
outside high-energy physics. Here, we systematically implement supersymmetric
quantum mechanics\textemdash exhibiting
fundamental SUSY properties in the simple setting of quantum mechanics\textemdash into
a wide range of topological semimetals, where the broken translational symmetry, e.g., by a magnetic field, is effectively captured by a SUSY potential. We show that the dynamical SUSY breaking via the instanton effect over the SUSY potential valleys works as the underlying mechanism for the gap opening of the topological semimetallic phases, and the magnitude of the instanton effect is proportional to the energy gap. This instanton mechanism provides a simple criterion for determining whether the energy gap has been opened, without resorting to detailed calculations, i.e., a finite energy gap is opened if and only if the SUSY potential has an even number of zeros. Our theory leads to previously unexpected results: even an infinitesimal magnetic field can open a gap in topologically robust Dirac, Weyl, and
nodal-line semimetallic phases due to the dynamical SUSY
breaking. Overall, the revealed connection between SUSY quantum mechanics and non-uniform topological semimetals can elucidate previously ambiguous phenomena, provide guidance for future investigations, and open a new avenue for exploring topological semimetals.
\end{abstract}
\maketitle

\textit{\textcolor{blue}{Introduction.}}\textit{\textemdash{}}
Supersymmetry (SUSY) attracts significant attention due to its profound potential for extending the Standard Model~\citep{gervais1971fieldtheory,golfand1971extension, volkov1998,dimopoulos1981softlybroken,witten1981dynamical,witten1982constraints,shadmi2000dynamical, nillesSupersymmetry1984,Haber-Supersymmetry-1985,martinSupersymmetryPrimer1998,baer2006weakscale,djouadi2008theanatomy}. To explain the absence of SUSY particles in current experiments, the spontaneous SUSY breaking at the low-energy regime is a necessary ingredient in SUSY quantum field theories~\citep{witten1982constraints,witten1981dynamical,shadmi2000dynamical}. To elucidate this core concept, Witten introduced supersymmetric quantum mechanics (SUSY QM)~\citep{witten1981dynamical}, namely the $(0+1)$D SUSY field theories, where the instanton effect acts as a key mechanism of dynamical SUSY breaking. Since then, it was soon clear that SUSY QM was interesting in its own right, and has been developed to understand various solvable potential problems~\citep{dutt-Supersymmetry-1988, arai-Supersymmetric1991,cooper1995supersymmetry,Gangopadhyaya2010}. In this work, we shall develop a new horizon of SUSY quantum mechanics, through revealing the connection between SUSY QM and  non-uniform topological semimetals. Before entering our topic, it would be noteworthy that SUSY theories have been realized or found applications in various fields outside of high-energy physics, such as optics, disordered systems, and condensed matters~\citep{junker1996supersymmetric,wegner2016supermathematics,efetov-Disorder-1996,Miri-prl-2013, heinrich2014supersymmetric,miri2013supersymmetric,garcia-meca2020supersymmetry,yu2010simulating,hokmabadi2019supersymmetric,ma2023gaugefield, yu2008supersymmetry,cai2022observation,minavr2022kinkdynamics, grover2014emergent,jian2015emergent,li2017edgequantum,weber2023secondorder,ponte2014emergence,yu2022emergent,ma2021realization, zhaiSupersymmetry2024,ezawa2023supersymmetric,prakash2021boundary,turzillo2021supersymmetric,behrends2020supersymmetry,nayga2019magnonlandau}. 
 
In condensed matter experiments, various non-uniform topological semimetals with translational symmetry breaking have been extensively studied. For instance, the translational symmetry may be violated under magnetic fields or strain, as actively being experimentally examined for topological Dirac, Weyl, and nodal-line semimetals~\citep{zhang-graphene-2005,li-weyl-2016, Huang-prx-2015, Zhao-prx-2015,zhao-weyl, shekhar-weyl-2015,wang3DQuantum2017, rui-weyl,yuanChirals2018a, kim-nodal-line-2018,zhang2017magnetictunnellinginduced,ramshaw2018quantum}. A central issue in these studies is to determine whether a finite critical field or strain is required for gap opening or the destruction of topological zero modes. Despite significant experimental and theoretical efforts, this question remains unresolved. Particularly, due to the lack of translational symmetry, conventional theoretical methods become ambiguous in capturing the topological essence of such non-uniform systems.

In this work, we establish a rigorous theoretical characterization for such non-uniform topological semimetals in terms of SUSY QM. We systematically implement SUSY QM into a wide range of non-uniform topological semimetals possessing chiral symmetry, where broken translational symmetry, e.g., by a magnetic field, is effectively captured by a SUSY potential. Remarkably, we uncover that the dynamical SUSY breaking, mediated by the instanton effect, serves as the mechanism underlying the gap opening or destruction of these semimetallic phases. The magnitude of the energy gap is proportional to instanton effect over the SUSY potential valleys, and can be determined by instanton calculations. Based on the instanton mechanism, we propose a straightforward criterion for determining whether an energy gap has been opened without requiring detailed calculations. Specifically, a finite energy gap is opened if and only if there are an even number of zeros in the SUSY potential.

Our theory yields several unexpected results. We find
that the Dirac points in graphene become gapped under a magnetic field,
even though the gap size is exponentially small. Furthermore, we demonstrate 
that pairs of Weyl points are destructed regardless of the strength of magnetic field  when oriented in a specific direction, clarifying that there is 
no threshold for the destruction to occur~\citep{zhang2017magnetictunnellinginduced,ramshaw2018quantum}.
Finally, we find that even a weak magnetic field can destroy
the topologically stable nodal-line semimetallic phases. These phenomena
all result from dynamical SUSY breaking via instanton effect, as identified by our proposed
method.

\textit{\textcolor{blue}{SUSY QM in topological phases possessing chiral symmetry.}}\textit{\textemdash{}}
Let us begin by examining topological materials characterized by a chiral-symmetric Hamiltonian denoted as $H$. Chiral symmetry $S$ demands that
$\{H,S\}=0$, where $S^{2}=\mathds{1}$ and $\mathcal{\mathds{1}}$
is the identity matrix. In the basis where $S$ is diagonal, the Hamiltonian
takes the block off-diagonal form. Thus, we can write $H$ as
\begin{equation}
H=\left(\begin{array}{cc}
0 & q\\
q^{\dagger} & 0
\end{array}\right)=\mathcal{Q}+\mathcal{Q}^{\dagger},\quad\mathcal{Q}=q\otimes\sigma_{+},\label{eq:chiral-hal}
\end{equation}
where $\sigma_{\pm}=1/2(\sigma_{1}\pm i\sigma_{2})$ and the $\sigma_{i}'s$
with $i=1,2,3$ are the Pauli matrices.Note that $q$ can be a matrix~\citep{Footnote-q}.

The connection of (\ref{eq:chiral-hal}) to SUSY QM is made by noting
that the operator $\mathcal{Q}$ satisfies
\begin{equation}
\mathcal{Q}^{2}=0,\label{eq: susy-algebra-1}
\end{equation}
which is identical to the algebra of supercharge in SUSY QM (See the
Supplementary Material (SM) for an introduction~\citep{SuppInf}).
By treating $\mathcal{Q}$ as the supercharge, a SUSY Hamiltonian
$\mathcal{H}$ can be constructed as
\begin{equation}
\mathcal{H}=\{\mathcal{Q},\mathcal{Q}^{\dagger}\}=H^{2},\label{eq:SUSY-hal}
\end{equation}
where $\{\mathcal{Q},\mathcal{Q}^{\dagger}\}=\mathcal{Q\mathcal{Q}^{\dagger}}+\mathcal{Q}^{\dagger}\mathcal{Q}$.
Equations~(\ref{eq: susy-algebra-1}) and (\ref{eq:SUSY-hal}) together
constitute the algebra of SUSY QM~\citep{junker1996supersymmetric},
confirming its emergence in the square of~(\ref{eq:chiral-hal}).

\textit{\textcolor{blue}{Correspondence between gapped/gapless topological phases and broken/unbroken SUSY.}}\textit{\textemdash{}}
From Eq.~(\ref{eq:SUSY-hal}),
eigenvalues $E$ of $H$ are linked to the SUSY eigenvalues $\mathcal{E}$
of $\mathcal{H}$ through the equation $E=\pm\sqrt{\mathcal{E}}$,
where the SUSY energy $\mathcal{E}$ is always non-negative. Hence, the
SUSY ground-state energy $\mathcal{E}_{g}$ is related to the band
gap $\Delta$ of the chiral-symmetric Hamiltonian $H$ as
\begin{equation}
\Delta=2\sqrt{\mathcal{E}_{g}}.\label{eq:gap-size}
\end{equation}
Notably, the ground-state energy determines whether SUSY is broken
or not: SUSY is \textit{broken} when the ground-state energy is positive
($\mathcal{E}_{g}>0$), and it is \textit{unbroken} when the ground-state
energy is zero ($\mathcal{E}_{g}=0$). As a result, we can associate
gapped topological phases (\textgreek{D} \ensuremath{\neq} 0) with
broken SUSY ($\mathcal{E}_{g}>0$), and gapless topological phases
(\textgreek{D} = 0) with unbroken SUSY ($\mathcal{E}_{g}=0$).

Moreover, broken or unbroken SUSY can be determined by the existence
of zero-energy ground state $|0\rangle$. From the energy expectation
value $\langle\mathcal{E}\rangle=\langle\psi|\mathcal{H}|\psi\rangle=(|\mathcal{Q}|\psi\rangle|^{2}+|\mathcal{Q}^{\dagger}|\psi\rangle|^{2})\geq0$,
it is evident that if $|0\rangle$ exists, it must satisfy $|\mathcal{Q}|0\rangle|^{2}=|\mathcal{Q}^{\dagger}|0\rangle|^{2}=0$,
namely, 
\begin{equation}
\mathcal{Q}|0\rangle=\mathcal{Q}^{\dagger}|0\rangle=0.\label{eq:zero-mode-hal}
\end{equation}
It means that the ground states of unbroken SUSY shall be annihilated
by supercharges.

\textit{\textcolor{blue}{Gap opening in topological semimetals by spontaneous SUSY breaking.}}\textit{\textemdash{}}
Spontaneous SUSY breaking happens from unbroken SUSY to broken SUSY.
From the above discussion, evidently, spontaneous SUSY breaking can trigger
novel gap-opening processes, which are crucial for topological materials. When applied to topological semimetallic phases, spontaneous SUSY breaking introduces new mechanisms that can destroy these phases.

\textit{\textcolor{blue}{Emergence of Witten's model in non-uniform topological
semimetals.}}\textit{\textemdash{} }
We are interested in the destruction
of topological semimetallic phases by spontaneous SUSY breaking mechanisms,
for which Witten's model~\citep{witten1981dynamical,junker1996supersymmetric} serves as a prominent and widely studied
example. We will thus focus on this model in our work. The supercharges
in Witten's model are defined by
\begin{align}
\mathcal{Q} & =\left[-i\partial_{x}-i\,w(x)\right]\otimes\sigma_{+},\label{eq:witten-charge}
\end{align}
and the corresponding SUSY Hamiltonian is given by
\begin{equation}
\mathcal{H}=\left[-\partial_{x}^{2}+w(x)^{2}\right]\otimes\mathds{1}+w'(x)\otimes\sigma_{3}.\label{eq:witten-model}
\end{equation}
Here, the SUSY potential $w(x)$ is an arbitrary real function of
the position $x$, and $w'(x)=\partial_{x}w(x)$. 

We find that Witten's model could emerge in topological semimetals
that are effectively characterized by the following $k\cdot p$ model,
\begin{equation}
Q(\mathbf{k})=[v_{i}k_{i}-ig(\tilde{\mathbf{k}})]\otimes\sigma_{+},\label{eq:kpmodel}
\end{equation}
where \(\mathbf{k} \in \mathbb{R}^d\) is the momentum in \(d\)-dimensional space, \(k_i\) is the component along direction \(i\) with \(i \in \{1, 2, \dots, d\}\), \(\tilde{\mathbf{k}} = (k_1, \dots, k_{i-1}, k_{i+1}, \dots, k_d)\) denotes the momentum excluding \(k_i\), \(v_{i}\) is a real parameter, and \(g(\tilde{\mathbf{k}})\) is a real function of \(\tilde{\mathbf{k}}\).
Notably, this
model is applicable to graphene
{[}Eqs.~(\ref{eq:graphene-single},\ref{eq:graphene-double}){]},
Weyl {[}Eq.~(\ref{eq:weyl}){]}, topological nodal-line {[}Eq.~(\ref{eq:hal-nodal-line}){]}
semimetals. Moreover, in the SM \citep{SuppInf}, we show that Eq.~\eqref{eq:kpmodel} can be generalized to systems with higher dimensions and systems with non-linear dispersion in $k_i$.

However, since Witten's model has no translational symmetry, external
factors such as magnetic fields or strain shall be applied to make 
the topological semimetals non-uniform. We choose external
factors that modify $g(\tilde{\mathbf{k}})$ to $g(x_{i},\tilde{\mathbf{k}})$,
making it to vary with respect to $x_{i}$ in real space, thereby
breaking the translational symmetry in the $i$-th direction. Hence, $Q(\mathbf{k})$
in~(\ref{eq:kpmodel}) is modified by,
\begin{equation}
v_{i}k_{i}\rightarrow-iv_{i}\partial_{i},\quad g(\tilde{\mathbf{k}})\rightarrow g(x_{i},\tilde{\mathbf{k}}),
\end{equation}
where $\partial_{i}=\partial_{x_i}$. By treating $g(x_{i},\tilde{\mathbf{k}})$ as the SUSY potential $w(x_i)$,
we arrive at the one-dimensional Witten's model~(\ref{eq:witten-charge}) in the direction $i$ up to a
prefactor $v_{i}$. It is worth noting that the rich variety of $g(\tilde{\mathbf{k}})$
in chiral-symmetric topological materials allows the realization of Witten's models with different types of SUSY potentials.

\textit{\textcolor{blue}{Criteria for determining gap opening.}}\textit{\textemdash{}} 
We find that the
novel gap-opening processes induced by spontaneous SUSY breaking can
be identified using Witten's criteria. In Witten's model,
assume the zero-energy state $|0\rangle$ takes a two-component form,
i.e., $|0\rangle=\left(\psi_{+}(x),\psi_{-}(x)\right)^{T}$ with $T$
the matrix transposition. Plug it into Eq.~(\ref{eq:zero-mode-hal}),
we find that $\left[-i\partial_{x}\pm iw(x)\right]\psi_{\pm}(x)=0$.
Solving these equations yields 
\begin{equation}
\psi_{\pm}(x)=e^{\pm\int_{0}^{x}dx'w(x')}.\label{eq:normalization condition}
\end{equation}
Hence, the zero-energy state exists if either $\psi_{+}(x)$ or $\psi_{-}(x)$
is normalizable. Note that we assume $|w(x)|\rightarrow\infty$ as
$|x|\rightarrow\infty$, so that the SUSY spectrum is discrete. 

\begin{figure}[t]
\includegraphics[width=0.9\columnwidth]{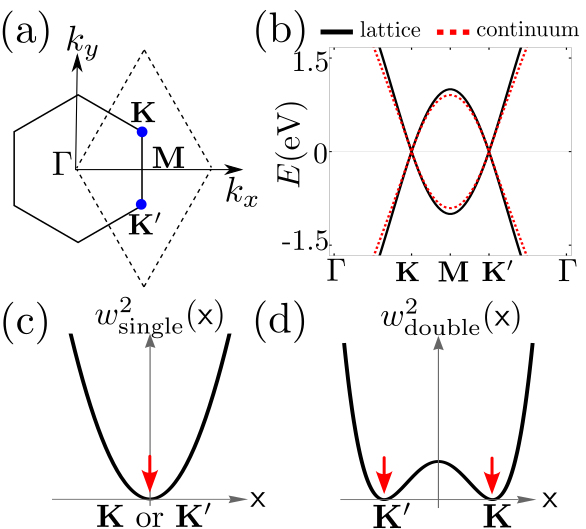}

\caption{(a) The momentum-space location of Dirac points in graphene. The dashed line indicates the reciprocal lattice unit cell.
(b) Comparison of spectrum between the tight-binding model~(\ref{eq:graphene-lattice})
and the $k\cdot p$ model for double Dirac points~(\ref{eq:graphene-double}).
(c) and (d) show the square of SUSY potential for single Dirac points
in Eq.~(\ref{eq:graphene-single}), and double Dirac points in Eq.~(\ref{eq: supercharge-single-dirac}).
Red arrows indicate the zeros of SUSY potential.\label{fig:graphene}}
\end{figure}

Witten established straightforward criteria for determining whether
SUSY is broken or not by merely counting the zeros of the SUSY potential
$w(x)$:
\begin{enumerate}
\item Unbroken SUSY: $w(x)$ has an \textit{odd number of zeros}. \\
In this case, the signs of $w(x)$ are opposite for $x\rightarrow+\infty$
and $x\rightarrow-\infty$, and one of $\psi_{\pm}(x)$ is normalizable. 
\item Broken SUSY: $w(x)$ has an \textit{even number of zeros}. \\
Here, the signs of $w(x)$ are the same for $x\rightarrow+\infty$
and $x\rightarrow-\infty$, and none of $\psi_{\pm}(x)$ is normalizable. 
\end{enumerate}
Notably, for non-zero even number of zeros, dynamical
SUSY breaking occurs due to the instanton effect over the SUSY potential 
valleys~\cite{witten1981dynamical}, as exemplified in Fig.~\ref{fig:graphene}(d). Therefore, gap opening 
can be determined by simply counting the zeros: 
a finite energy gap is opened if and only if the SUSY potential has an even number of zeros.

\textit{\textcolor{blue}{Gapped Dirac points in Graphene under a
magnetic field.}}\textit{\textemdash{}} 
We begin by considering the
tight-binding Hamiltonian for a monolayer graphene in momentum space,
given by~\citep{castroneto-Graphene-2009}
\begin{equation}
H_{g}(\mathbf{k})=t\left(\begin{array}{cc}
 & v(\mathbf{k})\\
v(\mathbf{k})^{*}
\end{array}\right),\label{eq:graphene-lattice}
\end{equation}
where $t\approx2.8\,$eV is the nearest-neighbor hopping amplitude,
$v(\mathbf{k})=1+e^{i\mathbf{k}\cdot\mathbf{a}_{1}}+e^{i\mathbf{k}\cdot\mathbf{a}_{2}}$,
and $\mathbf{a}_{1(2)}=a(3/2,\pm\sqrt{3}/2)$ are lattice vectors
with $a\approx1.42\,\text{Å}$. The Hamiltonian respects the chiral
symmetry $S=\sigma_{3}$. As shown Fig.~\ref{fig:graphene}(a), there
are two Dirac points located at $\mathbf{K}$ and $\mathbf{K'}$.

Previous research primarily uses the $k\cdot p$ model for single
Dirac points to study the effect of magnetic field. For example, the
$k\cdot p$ model for Dirac point at $\mathbf{K}$ reads
\begin{equation}
h_{\text{single}}(\mathbf{k})=v_{F}(k_{x}\sigma_{x}-k_{y}\sigma_{y}),\label{eq:graphene-single}
\end{equation}
where $v_{F}=3at/2$. SUSY QM can also be employed to study these
Hamiltonians~\citep{junker1996supersymmetric}. Consider a uniform magnetic field applied perpendicular
to the graphene sheet, i.e., $\mathbf{B}=(0,0,B)$, whose vector potential
is $\mathbf{A}=(0,Bx,0)$ under the Landau gauge. The potential enters
into the Hamiltonian via the replacement of $\mathbf{k}\rightarrow(-i\partial_{x},k_{y}+eBx/\hbar,k_{z}).$
We can further denote the replacement as
\begin{equation}
\mathbf{k}\rightarrow(-i\partial_{\mathsf{x}},\text{\ensuremath{\mathsf{x}}}/l_{B}^{2},k_{z}),\label{eq:magnetic-field}
\end{equation}
by performing a change of variable of $\mathsf{x}=x+k_{y}l_{B}^{2}$,
where $l_{B}=\sqrt{\hbar/eB}$ is the magnetic length. After the replacement,
the supercharge from Eq.~(\ref{eq:graphene-single}) is identified
as
\begin{equation}
\mathcal{Q}_{\text{single}}=\left[-iv_{F}\partial_{\mathsf{x}}-iw_{\text{single}}(\mathsf{x})\right]\otimes\sigma_{+}.\label{eq: supercharge-single-dirac}
\end{equation}
The square of SUSY potential $w_{\text{single}}(\mathsf{x})=-v_{F}/l_{B}^{2}\text{\ensuremath{\mathsf{x}}}$
is ploted in Fig.~\ref{fig:graphene}(c). Evidently, there is just
one zero for a single Dirac point. Hence, using Witten's
criteria, the spectrum is not gapped as SUSY is unbroken, in agreement
with previous studies. The full spectrum is derived in the SM~\citep{SuppInf}.

However, graphene possesses not one but two Dirac points, which means
that its physical properties cannot be adequately described by models
for single Dirac points like~(\ref{eq:graphene-single}). For this
reason, we construct a new $k\cdot p$ model capable of describing
both Dirac points as
\begin{equation}
H_{\text{double}}(\mathbf{k})=v_{x}k_{x}\sigma_{1}+\mu_{y}(K^{2}-k_{y}^{2})\sigma_{2},\label{eq:graphene-double}
\end{equation}
where $\mathbf{k}$ here is relative to the $\mathbf{M}$ point. Here,
$v_{x}=v_{F}$, $\mu_{y}=v_{F}/(2K)$, and $K=|\mathbf{K}-\mathbf{K'}|/2.$
This Hamiltonian reduces to those for single Dirac points in the vicinity of $\mathbf{K}$
or $\mathbf{K}'$, and agrees well with the lattice model of~(\ref{eq:graphene-lattice})
as compared in Fig.~\ref{fig:graphene}(b).

Similarly, the magnetic field enters into Eq.~(\ref{eq:graphene-double})
via the replacement of Eq.~(\ref{eq:magnetic-field}). We identify
the supercharge as
\begin{equation}
\mathcal{Q}_{\text{double}}=\left[-iv_{x}\partial_{\mathsf{x}}-iw_{\text{double}}(\mathsf{x})\right]\otimes\sigma_{+},\label{eq: supercharge-double-dirac}
\end{equation}
where $w_{\text{double}}(\mathsf{x})=\mu_{y}(K^{2}-\text{\ensuremath{\mathsf{x}}}^{2}/l_{B}^{4})$.
As shown in Fig.~\ref{fig:graphene}(d), the SUSY potential exhibits
two zeros in this case, due to the two Dirac points. Hence,
using the Witten's criteria, the spectrum is gapped due to spontaneous
SUSY breaking, in stark contrast with previous studies. We emphasize
that the two Dirac points are gapped no matter how weak the field
is. 

\textit{\textcolor{blue}{Gap opening due to dynamical SUSY breaking
via instanton effect.}}\textit{\textemdash{}} 
Remarkably, the gap
opening is due to dynamical SUSY breaking, namely,
very small quantum corrections that break SUSY at a low energy scale~\citep{witten1981dynamical,shadmi2000dynamical}. 

As shown in Fig.~\ref{fig:graphene}(d), the square of SUSY potential features two valleys around
the zeros, allowing for quantum tunneling between one and the
other, a phenomenon known as the instanton effect. It is this non-perturbative
effect that breaks SUSY and results in a non-zero ground-state energy.
The ground-state energy can be derived through the instanton calculation~\citep{marino2015instantons,vanholten1983instantons}(See
the SM for details~\citep{SuppInf}), from which the band gap is
determined as 
\begin{equation}
	\Delta=\Delta_{0} \sqrt{\lambda_0^3K/l_B^2}\exp{\left(-\frac{4}{3}\lambda_0 K^3l_B^2\right)}, \label{eq:dynamic-susy-breaking}
\end{equation}
where $\Delta_{0}=\sqrt{8v_x \mu_y/\pi\lambda_0^3}$ and $\lambda_0=\mu_y/v_x$.
Evidently, when $B\neq0$, no matter how small, the Dirac points
are gapped, aligning with Witten's criteria. We find that due to
the large separation of Dirac points ($K=8.5\,$nm$^{-1}$), the gap
size is exponentially small, e.g., $\Delta=10$ meV for $B=10^{4}\, \text{T}$.

\begin{figure}[t]
	\includegraphics[width=0.96\columnwidth]{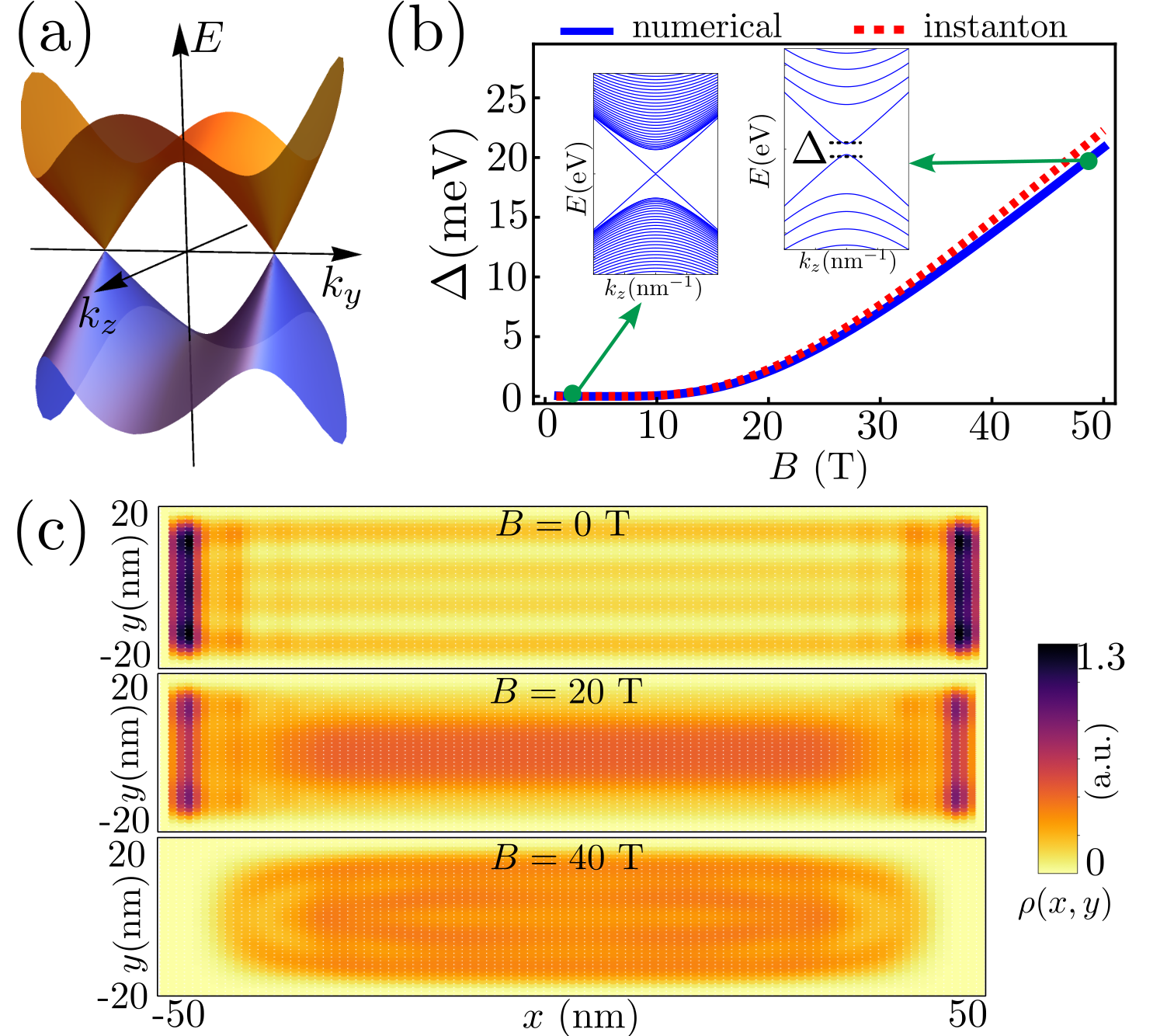}
	
	\caption{(a) The pair of Weyl points along the $y$-direction. (b) The band
		gap obtained by the numerical simulation (blue) using a continuum method~\citep{SuppInf} and the analytical instanton calculation
		(red). The insets show two representative energy spectra. (c) LDOS
		of $\rho(x,y)$ at $E=0$ under three different magnetic fields. The sample is periodic 
		in the $z$-drection and has dimensions of
		$L=100\,$nm and $W=40\,$nm in the $x$- and $y$-directions, respectively.
		The parameters are
		$v_{x}=0.2\,$eV$\cdot$nm, $\mu_{y}=0.5\,$eV$\cdot$nm$^{2}$,
		$v_{z}=0.1\,$eV$\cdot$nm, and $K=0.3\,$nm$^{-1}$. \label{fig:weyl}}
\end{figure}

\textit{\textcolor{blue}{Gap opening in Weyl semimetals under a weak magnetic field.}}\textit{\textemdash{}} 
In experiments~\citep{zhang2017magnetictunnellinginduced,ramshaw2018quantum},
it was observed that pairs of Weyl points with opposite chiralities
are gapped when a magnetic field is applied perpendicular to
their separation direction. Furthermore, these studies suggested that
the field strength must exceed a certain threshold to observe the
gap opening. Here, we clarify that no such threshold exists for
gap opening: it occurs as long as the magnetic
field is present, due to dynamical SUSY breaking. 

Let us consider a pair of Weyl points separated along the $y$-direction,
as shown in Fig.~\ref{fig:weyl}(a), with the Hamiltonian given by
\begin{equation}
	H_{\text{w}}(\mathbf{k})=v_{x}k_{x}\sigma_{1}+\mu_{y}(K^{2}-k_{y}^{2})\sigma_{2}+v_{z}k_{z}\sigma_{3},\label{eq:weyl}
\end{equation}
where the separation $K$ can be small in Weyl semimetals~\citep{zhang2017magnetictunnellinginduced,ramshaw2018quantum}. 
To determine the gap opening upon applying a perpendicular magnetic field (i.e., in the $z$-direction), it suffices 
to consider the Hamiltonian at  $k_{z}=0$, as the projected bands for $k_z\neq 0$ already possess a gap~\citep{footnote}.
This brings us back to
the same scenario for the double Dirac points, since the Hamiltonian at $k_z=0$
reduces to Eq.~(\ref{eq:graphene-double}) with chiral symmetry.
Given the two zeros in the SUSY potential~{[}Fig.~\ref{fig:graphene}(d){]},
the presence of a magnetic field, regardless of its strength, results
in the gap opening in Weyl semimetals due to instanton effect. 

As shown in Fig.~\ref{fig:weyl}(b), the gap size 
is extremely small for field strength below a certain value due to
the nature of the exponential function, consistent with the
experimental observation of a threshold~\citep{Footnote3}. However, our analytical instanton calculation
of Eq.~(\ref{eq:dynamic-susy-breaking}) unambiguously clarify that
as long as the magnetic field exists, the gap opens. Our results are further confirmed by numerical
calculations based on the continuum model of~(\ref{eq:weyl}),
as shown in Fig.~\ref{fig:weyl}(b)~\citep{SuppInf}.

\begin{figure}[t]
	\includegraphics[width=0.9\columnwidth]{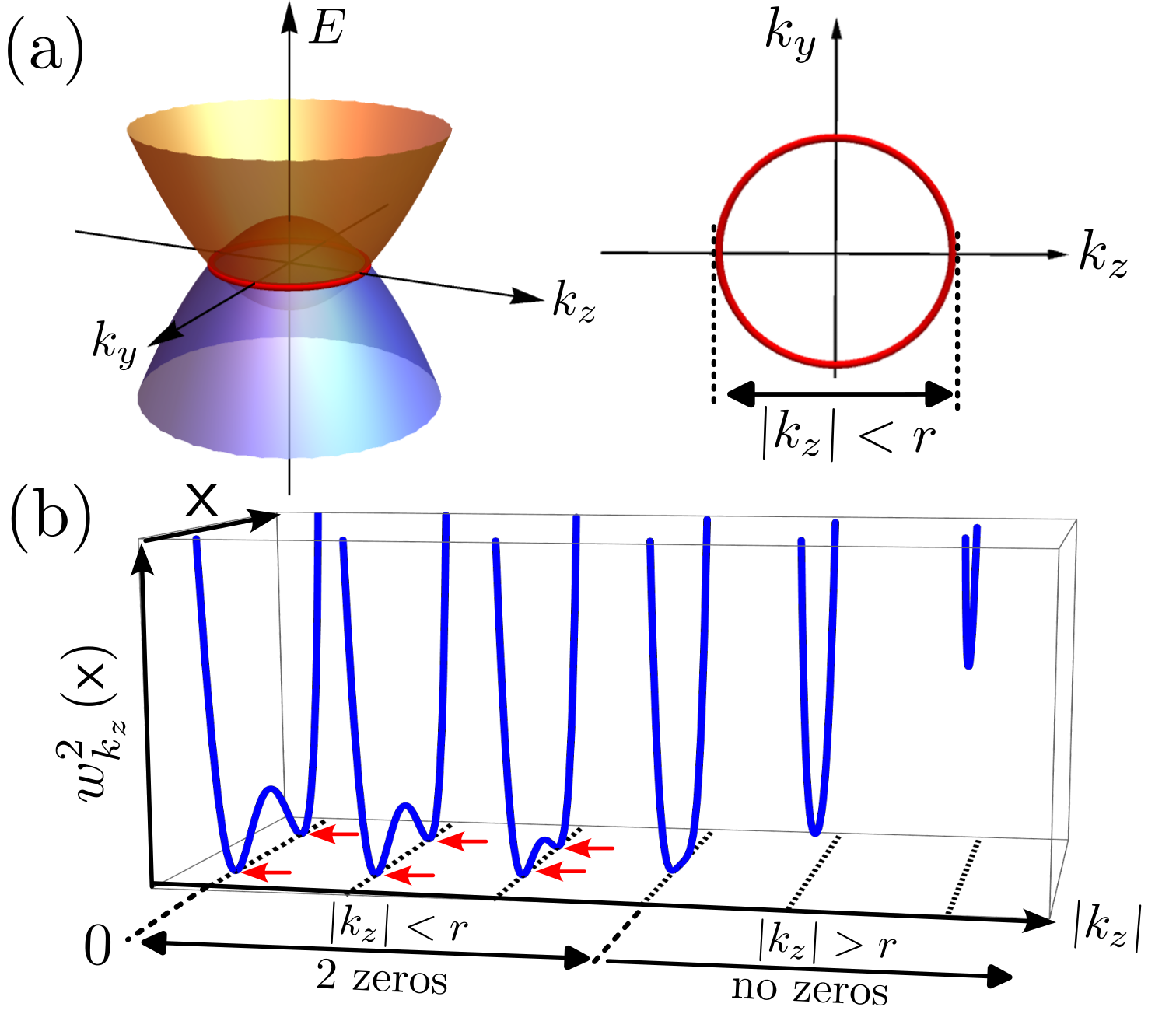}
	
	\caption{(a) Spectrum of the topological nodal-line semimetal at $k_{x}=0$
		(left). The noal line is located on the $(k_{y},k_{z})$-plane (right).
		(b) The squared SUSY potential $w_{k_{z}}^{2}(\mathsf{x})$ for different
		$k_{z}$ values. Red arrows highlight the zeros. \label{fig:Fig2}}
\end{figure}

\textit{\textcolor{blue}{Experimental proposal.}}\textit{\textemdash{}} 
We propose scanning tunneling spectroscopy (STS) as a direct method to investigate the gap opening. STS, well recognized for its ability to measure Landau levels in topological semimetals~\citep{liLandauLevels2007,jeon2014}, can directly observe the characteristic scaling of Eq.~\eqref{eq:dynamic-susy-breaking} against the magnetic field strength, thereby confirming the instanton-induced SUSY breaking mechanism.

Additionally, the gap opening influences topological properties, manifesting as observable changes. As shown in Fig.~\ref{fig:weyl}(c), we compute
the local density of states (LDOS) at zero energy under different
magnetic fields in Weyl semimetals. At $B=0\, \text{T}$, the Fermi-arc surface states are localized
at the two boundaries in the $x$-direction. As the magnetic field
increases to $B=20\, \text{T}$, a noticeable reduction in the boundary LDOS
is observed, and at $B=40 \,\text{T}$, the localization is  strongly suppressed.
Therefore, the destruction of Weyl points is also detectable by LDOS, which can be 
observed in STM experiments~\citep{schwenk-STM-2020}.

\textit{\textcolor{blue}{Destruction of topological nodal-line semimetallic
phase by spontaneous SUSY breaking.}}\textit{\textemdash{}} 
Moreover,
we apply our theory to topological nodal-line semimetals, with a typical
$k\cdot p$ model given by~\citep{chen-nl-2015,Li-nodal-line,rui2018topological}
\begin{equation}
H_{\text{nl}}(\mathbf{k})=v_{x}k_{x}\sigma_{1}+t\left[r^{2}-(k_{y}^{2}+k_{z}^{2})\right]\sigma_{2},\label{eq:hal-nodal-line}
\end{equation}
 which has the chiral symmetry $S=\sigma_{3}$. Here, $v_{x}$, $t$,
and $r$ are model parameters. As shown in Fig.~\ref{fig:Fig2}(a),
the nodal points form a ring with radius $r$. 

Upon applying a magnetic field in the $z$-direction, the supercharge
is obtained by the replacement
of Eq.~(\ref{eq:magnetic-field}) in $H_{\text{nl}}$, namely,
\begin{equation}
\mathcal{Q}_{\text{nl}}=\left[-iv_{x}\partial_{\mathsf{x}}-iw_{k_{z}}(\mathsf{x})\right]\otimes\sigma_{+},\label{eq:supercharge-nl}
\end{equation}
 where $w_{k_{z}}(\mathsf{x})=t\left[r^{2}(k_{z})-\mathsf{x}^{2}/l_{B}^{4}\right]$
and $r^{2}(k_{z})=r^{2}-k_{z}^{2}$. Unlike previous cases, the SUSY
potential $w_{k_{z}}(\mathsf{x})$ now depends on $k_{z}$. As shown
in Fig.~\ref{fig:Fig2}(b), within the nodal ring ($|k_{z}|<r$), the SUSY potential exhibits
two zeros, whereas no zeros are found outside the ring ($|k_{z}|>r$).
Therefore, using the Witten's criteria, the spectrum is gapped due to the instanton effect. The calculation of the full spectrum can be found in the SM~\citep{SuppInf}.

\textit{\textcolor{blue}{Discussions.}}\textit{\textemdash{}} 
We have uncovered SUSY QM in a wide range of non-uniform topological semimetals possessing chiral symmetry, and revealed the underlying mechanism of a novel kind of gap opening in these materials as dynamical SUSY breaking via the instanton effect.
Besides the rich variety of topological materials with chiral symmetry~\citep{Chiu_RMP_2016}, our theory could also find applications in systems where chiral-symmetric models serve as effective descriptions~\citep{chiral-limit-tbg,Chiral-approximation-tbg}. In the SM~\citep{SuppInf}, we have further generalized our theoretical framework to include discussions of higher-order SUSY QM~\citep{HigherorderSupersymmetricQuantum2004} and to investigate the higher-dimensional extensions.

Finally, we note that the SUSY uncovered in this work operates within the framework of SUSY QM, exhibiting fundamental SUSY properties such as Bose-Fermi degeneracy and broken/unbroken SUSY in a simple yet experimentally accessible setting.
It would be interesting to investigate if SUSY can similarly arise in more complex contexts, such as SUSY quantum field
theories pursued by particle physicists, within condensed matter physics.

\vspace{0.5cm}
\begin{acknowledgments}
	W.B.R. is grateful to Shaojie Ma, Dawei Zhai, Chenjie Wang, and Bo Fu for valuable discussions.
	This work was supported by the Guangdong-Hong Kong Joint Laboratory of Quantum Matter, the Quantum Science Center of Guangdong-Hong Kong-Macau Greater Bay Area, the NSFC/RGC JRS grant (RGC Grant No. N\_HKU774/21, NSFC Grant No. 12161160315), and the GRF of Hong Kong (Grants No. 17310622, No. 17303023, and No. 17301224). W.B.R. was supported by the RGC Postdoctoral Fellowship (Ref. No. PDFS2223-7S05).
\end{acknowledgments}

\bibliography{Reference}

\begin{thebibliography}{72}%
\makeatletter
\providecommand \@ifxundefined [1]{%
 \@ifx{#1\undefined}
}%
\providecommand \@ifnum [1]{%
 \ifnum #1\expandafter \@firstoftwo
 \else \expandafter \@secondoftwo
 \fi
}%
\providecommand \@ifx [1]{%
 \ifx #1\expandafter \@firstoftwo
 \else \expandafter \@secondoftwo
 \fi
}%
\providecommand \natexlab [1]{#1}%
\providecommand \enquote  [1]{``#1''}%
\providecommand \bibnamefont  [1]{#1}%
\providecommand \bibfnamefont [1]{#1}%
\providecommand \citenamefont [1]{#1}%
\providecommand \href@noop [0]{\@secondoftwo}%
\providecommand \href [0]{\begingroup \@sanitize@url \@href}%
\providecommand \@href[1]{\@@startlink{#1}\@@href}%
\providecommand \@@href[1]{\endgroup#1\@@endlink}%
\providecommand \@sanitize@url [0]{\catcode `\\12\catcode `\$12\catcode
  `\&12\catcode `\#12\catcode `\^12\catcode `\_12\catcode `\%12\relax}%
\providecommand \@@startlink[1]{}%
\providecommand \@@endlink[0]{}%
\providecommand \url  [0]{\begingroup\@sanitize@url \@url }%
\providecommand \@url [1]{\endgroup\@href {#1}{\urlprefix }}%
\providecommand \urlprefix  [0]{URL }%
\providecommand \Eprint [0]{\href }%
\providecommand \doibase [0]{https://doi.org/}%
\providecommand \selectlanguage [0]{\@gobble}%
\providecommand \bibinfo  [0]{\@secondoftwo}%
\providecommand \bibfield  [0]{\@secondoftwo}%
\providecommand \translation [1]{[#1]}%
\providecommand \BibitemOpen [0]{}%
\providecommand \bibitemStop [0]{}%
\providecommand \bibitemNoStop [0]{.\EOS\space}%
\providecommand \EOS [0]{\spacefactor3000\relax}%
\providecommand \BibitemShut  [1]{\csname bibitem#1\endcsname}%
\let\auto@bib@innerbib\@empty
\bibitem [{\citenamefont {Gervais}\ and\ \citenamefont
  {Sakita}(1971)}]{gervais1971fieldtheory}%
  \BibitemOpen
  \bibfield  {author} {\bibinfo {author} {\bibfnamefont {J.~L.}\ \bibnamefont
  {Gervais}}\ and\ \bibinfo {author} {\bibfnamefont {B.}~\bibnamefont
  {Sakita}},\ }\bibfield  {title} {\bibinfo {title} {Field theory
  interpretation of supergauges in dual models},\ }\href
  {https://doi.org/10.1016/0550-3213(71)90351-8} {\bibfield  {journal}
  {\bibinfo  {journal} {Nuclear Physics B}\ }\textbf {\bibinfo {volume} {34}},\
  \bibinfo {pages} {632} (\bibinfo {year} {1971})}\BibitemShut {NoStop}%
\bibitem [{\citenamefont {Golfand}\ and\ \citenamefont
  {Likhtman}(1971)}]{golfand1971extension}%
  \BibitemOpen
  \bibfield  {author} {\bibinfo {author} {\bibfnamefont {Y.~A.}\ \bibnamefont
  {Golfand}}\ and\ \bibinfo {author} {\bibfnamefont {E.~P.}\ \bibnamefont
  {Likhtman}},\ }\bibfield  {title} {\bibinfo {title} {Extension of the algebra
  of poincare group generators and violation of p invariance},\ }\href@noop {}
  {\bibfield  {journal} {\bibinfo  {journal} {JETP Letters [translation of
  Pisma v Zhurnal Eksperimentalnoi i Teoreticheskoi Fiziki]}\ }\textbf
  {\bibinfo {volume} {13}},\ \bibinfo {pages} {323} (\bibinfo {year}
  {1971})}\BibitemShut {NoStop}%
\bibitem [{\citenamefont {Volkov}\ and\ \citenamefont
  {Akulov}(1998)}]{volkov1998}%
  \BibitemOpen
  \bibfield  {author} {\bibinfo {author} {\bibfnamefont {D.~V.}\ \bibnamefont
  {Volkov}}\ and\ \bibinfo {author} {\bibfnamefont {V.~P.}\ \bibnamefont
  {Akulov}},\ }\bibfield  {title} {\bibinfo {title} {Possible universal
  neutrino interaction},\ }in\ \href {https://doi.org/10.1007/BFb0105270}
  {\emph {\bibinfo {booktitle} {Supersymmetry and {{Quantum Field Theory}}}}},\
  \bibinfo {editor} {edited by\ \bibinfo {editor} {\bibfnamefont
  {J.}~\bibnamefont {Wess}}\ and\ \bibinfo {editor} {\bibfnamefont {V.~P.}\
  \bibnamefont {Akulov}}}\ (\bibinfo  {publisher} {Springer},\ \bibinfo
  {address} {Berlin, Heidelberg},\ \bibinfo {year} {1998})\ pp.\ \bibinfo
  {pages} {383--385}\BibitemShut {NoStop}%
\bibitem [{\citenamefont {Dimopoulos}\ and\ \citenamefont
  {Georgi}(1981)}]{dimopoulos1981softlybroken}%
  \BibitemOpen
  \bibfield  {author} {\bibinfo {author} {\bibfnamefont {S.}~\bibnamefont
  {Dimopoulos}}\ and\ \bibinfo {author} {\bibfnamefont {H.}~\bibnamefont
  {Georgi}},\ }\bibfield  {title} {\bibinfo {title} {Softly broken
  supersymmetry and {SU}(5)},\ }\href
  {https://doi.org/10.1016/0550-3213(81)90522-8} {\bibfield  {journal}
  {\bibinfo  {journal} {Nuclear Physics B}\ }\textbf {\bibinfo {volume}
  {193}},\ \bibinfo {pages} {150} (\bibinfo {year} {1981})}\BibitemShut
  {NoStop}%
\bibitem [{\citenamefont {Witten}(1981)}]{witten1981dynamical}%
  \BibitemOpen
  \bibfield  {author} {\bibinfo {author} {\bibfnamefont {E.}~\bibnamefont
  {Witten}},\ }\bibfield  {title} {\bibinfo {title} {Dynamical breaking of
  supersymmetry},\ }\href {https://doi.org/10.1016/0550-3213(81)90006-7}
  {\bibfield  {journal} {\bibinfo  {journal} {Nuclear Physics B}\ }\textbf
  {\bibinfo {volume} {188}},\ \bibinfo {pages} {513} (\bibinfo {year}
  {1981})}\BibitemShut {NoStop}%
\bibitem [{\citenamefont {Witten}(1982)}]{witten1982constraints}%
  \BibitemOpen
  \bibfield  {author} {\bibinfo {author} {\bibfnamefont {E.}~\bibnamefont
  {Witten}},\ }\bibfield  {title} {\bibinfo {title} {Constraints on
  supersymmetry breaking},\ }\href
  {https://doi.org/10.1016/0550-3213(82)90071-2} {\bibfield  {journal}
  {\bibinfo  {journal} {Nuclear Physics B}\ }\textbf {\bibinfo {volume}
  {202}},\ \bibinfo {pages} {253} (\bibinfo {year} {1982})}\BibitemShut
  {NoStop}%
\bibitem [{\citenamefont {Shadmi}\ and\ \citenamefont
  {Shirman}(2000)}]{shadmi2000dynamical}%
  \BibitemOpen
  \bibfield  {author} {\bibinfo {author} {\bibfnamefont {Y.}~\bibnamefont
  {Shadmi}}\ and\ \bibinfo {author} {\bibfnamefont {Y.}~\bibnamefont
  {Shirman}},\ }\bibfield  {title} {\bibinfo {title} {Dynamical supersymmetry
  breaking},\ }\href {https://doi.org/10.1103/RevModPhys.72.25} {\bibfield
  {journal} {\bibinfo  {journal} {Reviews of Modern Physics}\ }\textbf
  {\bibinfo {volume} {72}},\ \bibinfo {pages} {25} (\bibinfo {year}
  {2000})}\BibitemShut {NoStop}%
\bibitem [{\citenamefont {Nilles}(1984)}]{nillesSupersymmetry1984}%
  \BibitemOpen
  \bibfield  {author} {\bibinfo {author} {\bibfnamefont {H.~P.}\ \bibnamefont
  {Nilles}},\ }\bibfield  {title} {\bibinfo {title} {Supersymmetry,
  supergravity and particle physics},\ }\href
  {https://doi.org/10.1016/0370-1573(84)90008-5} {\bibfield  {journal}
  {\bibinfo  {journal} {Physics Reports}\ }\textbf {\bibinfo {volume} {110}},\
  \bibinfo {pages} {1} (\bibinfo {year} {1984})}\BibitemShut {NoStop}%
\bibitem [{\citenamefont {Haber}\ and\ \citenamefont
  {Kane}(1985)}]{Haber-Supersymmetry-1985}%
  \BibitemOpen
  \bibfield  {author} {\bibinfo {author} {\bibfnamefont {H.~E.}\ \bibnamefont
  {Haber}}\ and\ \bibinfo {author} {\bibfnamefont {G.~L.}\ \bibnamefont
  {Kane}},\ }\bibfield  {title} {\bibinfo {title} {The search for
  supersymmetry: {{Probing}} physics beyond the standard model},\ }\href
  {https://doi.org/10.1016/0370-1573(85)90051-1} {\bibfield  {journal}
  {\bibinfo  {journal} {Physics Reports}\ }\textbf {\bibinfo {volume} {117}},\
  \bibinfo {pages} {75} (\bibinfo {year} {1985})}\BibitemShut {NoStop}%
\bibitem [{\citenamefont {Martin}(1998)}]{martinSupersymmetryPrimer1998}%
  \BibitemOpen
  \bibfield  {author} {\bibinfo {author} {\bibfnamefont {S.~P.}\ \bibnamefont
  {Martin}},\ }\bibfield  {title} {\bibinfo {title} {A supersymmetry primer},\
  }in\ \href {https://doi.org/10.1142/9789812839657_0001} {\emph {\bibinfo
  {booktitle} {Perspectives on {{Supersymmetry}}}}},\ \bibinfo {series}
  {Advanced {{Series}} on {{Directions}} in {{High Energy Physics}}}, Vol.\
  \bibinfo {volume} {Volume 18}\ (\bibinfo  {publisher} {WORLD SCIENTIFIC},\
  \bibinfo {year} {1998})\ pp.\ \bibinfo {pages} {1--98}\BibitemShut {NoStop}%
\bibitem [{\citenamefont {Baer}\ and\ \citenamefont
  {Tata}(2006)}]{baer2006weakscale}%
  \BibitemOpen
  \bibfield  {author} {\bibinfo {author} {\bibfnamefont {H.}~\bibnamefont
  {Baer}}\ and\ \bibinfo {author} {\bibfnamefont {X.}~\bibnamefont {Tata}},\
  }\href {https://doi.org/10.1017/CBO9780511617270} {\emph {\bibinfo {title}
  {Weak {Scale} {Supersymmetry}: {From} {Superfields} to {Scattering}
  {Events}}}}\ (\bibinfo  {publisher} {Cambridge University Press},\ \bibinfo
  {address} {Cambridge},\ \bibinfo {year} {2006})\BibitemShut {NoStop}%
\bibitem [{\citenamefont {Djouadi}(2008)}]{djouadi2008theanatomy}%
  \BibitemOpen
  \bibfield  {author} {\bibinfo {author} {\bibfnamefont {A.}~\bibnamefont
  {Djouadi}},\ }\bibfield  {title} {\bibinfo {title} {The anatomy of
  electroweak symmetry breaking {Tome} {II}: {The} {Higgs} bosons in the
  {Minimal} {Supersymmetric} {Model}},\ }\href
  {https://doi.org/10.1016/j.physrep.2007.10.005} {\bibfield  {journal}
  {\bibinfo  {journal} {Physics Reports}\ }\textbf {\bibinfo {volume} {459}},\
  \bibinfo {pages} {1} (\bibinfo {year} {2008})}\BibitemShut {NoStop}%
\bibitem [{\citenamefont {Dutt}\ \emph {et~al.}(1988)\citenamefont {Dutt},
  \citenamefont {Khare},\ and\ \citenamefont
  {Sukhatme}}]{dutt-Supersymmetry-1988}%
  \BibitemOpen
  \bibfield  {author} {\bibinfo {author} {\bibfnamefont {R.}~\bibnamefont
  {Dutt}}, \bibinfo {author} {\bibfnamefont {A.}~\bibnamefont {Khare}},\ and\
  \bibinfo {author} {\bibfnamefont {U.~P.}\ \bibnamefont {Sukhatme}},\
  }\bibfield  {title} {\bibinfo {title} {Supersymmetry, shape invariance, and
  exactly solvable potentials},\ }\href {https://doi.org/10.1119/1.15697}
  {\bibfield  {journal} {\bibinfo  {journal} {American Journal of Physics}\
  }\textbf {\bibinfo {volume} {56}},\ \bibinfo {pages} {163} (\bibinfo {year}
  {1988})}\BibitemShut {NoStop}%
\bibitem [{\citenamefont {Arai}(1991)}]{arai-Supersymmetric1991}%
  \BibitemOpen
  \bibfield  {author} {\bibinfo {author} {\bibfnamefont {A.}~\bibnamefont
  {Arai}},\ }\bibfield  {title} {\bibinfo {title} {Exactly solvable
  supersymmetric quantum mechanics},\ }\href
  {https://doi.org/10.1016/0022-247X(91)90267-4} {\bibfield  {journal}
  {\bibinfo  {journal} {Journal of Mathematical Analysis and Applications}\
  }\textbf {\bibinfo {volume} {158}},\ \bibinfo {pages} {63} (\bibinfo {year}
  {1991})}\BibitemShut {NoStop}%
\bibitem [{\citenamefont {Cooper}\ \emph {et~al.}(1995)\citenamefont {Cooper},
  \citenamefont {Khare},\ and\ \citenamefont
  {Sukhatme}}]{cooper1995supersymmetry}%
  \BibitemOpen
  \bibfield  {author} {\bibinfo {author} {\bibfnamefont {F.}~\bibnamefont
  {Cooper}}, \bibinfo {author} {\bibfnamefont {A.}~\bibnamefont {Khare}},\ and\
  \bibinfo {author} {\bibfnamefont {U.}~\bibnamefont {Sukhatme}},\ }\bibfield
  {title} {\bibinfo {title} {Supersymmetry and quantum mechanics},\ }\href
  {https://doi.org/10.1016/0370-1573(94)00080-M} {\bibfield  {journal}
  {\bibinfo  {journal} {Physics Reports}\ }\textbf {\bibinfo {volume} {251}},\
  \bibinfo {pages} {267} (\bibinfo {year} {1995})}\BibitemShut {NoStop}%
\bibitem [{\citenamefont {Gangopadhyaya}\ \emph {et~al.}(2010)\citenamefont
  {Gangopadhyaya}, \citenamefont {Mallow},\ and\ \citenamefont
  {Rasinariu}}]{Gangopadhyaya2010}%
  \BibitemOpen
  \bibfield  {author} {\bibinfo {author} {\bibfnamefont {A.}~\bibnamefont
  {Gangopadhyaya}}, \bibinfo {author} {\bibfnamefont {J.~V.}\ \bibnamefont
  {Mallow}},\ and\ \bibinfo {author} {\bibfnamefont {C.}~\bibnamefont
  {Rasinariu}},\ }\href {https://doi.org/10.1142/7788} {\emph {\bibinfo {title}
  {Supersymmetric Quantum Mechanics}}}\ (\bibinfo  {publisher} {WORLD
  SCIENTIFIC},\ \bibinfo {year} {2010})\BibitemShut {NoStop}%
\bibitem [{\citenamefont {Junker}(1996)}]{junker1996supersymmetric}%
  \BibitemOpen
  \bibfield  {author} {\bibinfo {author} {\bibfnamefont {G.}~\bibnamefont
  {Junker}},\ }\href {https://doi.org/10.1007/978-3-642-61194-0} {\emph
  {\bibinfo {title} {Supersymmetric {Methods} in {Quantum} and {Statistical}
  {Physics}}}}\ (\bibinfo  {publisher} {Springer},\ \bibinfo {address} {Berlin,
  Heidelberg},\ \bibinfo {year} {1996})\BibitemShut {NoStop}%
\bibitem [{\citenamefont {Wegner}(2016)}]{wegner2016supermathematics}%
  \BibitemOpen
  \bibfield  {author} {\bibinfo {author} {\bibfnamefont {F.}~\bibnamefont
  {Wegner}},\ }\href {https://doi.org/10.1007/978-3-662-49170-6} {\emph
  {\bibinfo {title} {Supermathematics and its {Applications} in {Statistical}
  {Physics}}}},\ \bibinfo {series} {Lecture {Notes} in {Physics}}, Vol.\
  \bibinfo {volume} {920}\ (\bibinfo  {publisher} {Springer},\ \bibinfo
  {address} {Berlin, Heidelberg},\ \bibinfo {year} {2016})\BibitemShut
  {NoStop}%
\bibitem [{\citenamefont {Efetov}(1996)}]{efetov-Disorder-1996}%
  \BibitemOpen
  \bibfield  {author} {\bibinfo {author} {\bibfnamefont {K.}~\bibnamefont
  {Efetov}},\ }\href {https://doi.org/10.1017/CBO9780511573057} {\emph
  {\bibinfo {title} {Supersymmetry in {{Disorder}} and {{Chaos}}}}}\ (\bibinfo
  {publisher} {Cambridge University Press},\ \bibinfo {address} {Cambridge},\
  \bibinfo {year} {1996})\BibitemShut {NoStop}%
\bibitem [{\citenamefont {Miri}\ \emph
  {et~al.}(2013{\natexlab{a}})\citenamefont {Miri}, \citenamefont {Heinrich},
  \citenamefont {El-Ganainy},\ and\ \citenamefont
  {Christodoulides}}]{Miri-prl-2013}%
  \BibitemOpen
  \bibfield  {author} {\bibinfo {author} {\bibfnamefont {M.-A.}\ \bibnamefont
  {Miri}}, \bibinfo {author} {\bibfnamefont {M.}~\bibnamefont {Heinrich}},
  \bibinfo {author} {\bibfnamefont {R.}~\bibnamefont {El-Ganainy}},\ and\
  \bibinfo {author} {\bibfnamefont {D.~N.}\ \bibnamefont {Christodoulides}},\
  }\bibfield  {title} {\bibinfo {title} {Supersymmetric optical structures},\
  }\href {https://doi.org/10.1103/PhysRevLett.110.233902} {\bibfield  {journal}
  {\bibinfo  {journal} {Phys. Rev. Lett.}\ }\textbf {\bibinfo {volume} {110}},\
  \bibinfo {pages} {233902} (\bibinfo {year} {2013}{\natexlab{a}})}\BibitemShut
  {NoStop}%
\bibitem [{\citenamefont {Heinrich}\ \emph {et~al.}(2014)\citenamefont
  {Heinrich}, \citenamefont {Miri}, \citenamefont {St{\"u}tzer}, \citenamefont
  {El-Ganainy}, \citenamefont {Nolte}, \citenamefont {Szameit},\ and\
  \citenamefont {Christodoulides}}]{heinrich2014supersymmetric}%
  \BibitemOpen
  \bibfield  {author} {\bibinfo {author} {\bibfnamefont {M.}~\bibnamefont
  {Heinrich}}, \bibinfo {author} {\bibfnamefont {M.-A.}\ \bibnamefont {Miri}},
  \bibinfo {author} {\bibfnamefont {S.}~\bibnamefont {St{\"u}tzer}}, \bibinfo
  {author} {\bibfnamefont {R.}~\bibnamefont {El-Ganainy}}, \bibinfo {author}
  {\bibfnamefont {S.}~\bibnamefont {Nolte}}, \bibinfo {author} {\bibfnamefont
  {A.}~\bibnamefont {Szameit}},\ and\ \bibinfo {author} {\bibfnamefont {D.~N.}\
  \bibnamefont {Christodoulides}},\ }\bibfield  {title} {\bibinfo {title}
  {Supersymmetric mode converters},\ }\href
  {https://doi.org/10.1038/ncomms4698} {\bibfield  {journal} {\bibinfo
  {journal} {Nature Communications}\ }\textbf {\bibinfo {volume} {5}},\
  \bibinfo {pages} {3698} (\bibinfo {year} {2014})}\BibitemShut {NoStop}%
\bibitem [{\citenamefont {Miri}\ \emph
  {et~al.}(2013{\natexlab{b}})\citenamefont {Miri}, \citenamefont {Heinrich},
  \citenamefont {El-Ganainy},\ and\ \citenamefont
  {Christodoulides}}]{miri2013supersymmetric}%
  \BibitemOpen
  \bibfield  {author} {\bibinfo {author} {\bibfnamefont {M.-A.}\ \bibnamefont
  {Miri}}, \bibinfo {author} {\bibfnamefont {M.}~\bibnamefont {Heinrich}},
  \bibinfo {author} {\bibfnamefont {R.}~\bibnamefont {El-Ganainy}},\ and\
  \bibinfo {author} {\bibfnamefont {D.~N.}\ \bibnamefont {Christodoulides}},\
  }\bibfield  {title} {\bibinfo {title} {Supersymmetric {Optical}
  {Structures}},\ }\href {https://doi.org/10.1103/PhysRevLett.110.233902}
  {\bibfield  {journal} {\bibinfo  {journal} {Physical Review Letters}\
  }\textbf {\bibinfo {volume} {110}},\ \bibinfo {pages} {233902} (\bibinfo
  {year} {2013}{\natexlab{b}})}\BibitemShut {NoStop}%
\bibitem [{\citenamefont {Garc{\'i}a-Meca}\ \emph {et~al.}(2020)\citenamefont
  {Garc{\'i}a-Meca}, \citenamefont {Ortiz},\ and\ \citenamefont
  {S{\'a}ez}}]{garcia-meca2020supersymmetry}%
  \BibitemOpen
  \bibfield  {author} {\bibinfo {author} {\bibfnamefont {C.}~\bibnamefont
  {Garc{\'i}a-Meca}}, \bibinfo {author} {\bibfnamefont {A.~M.}\ \bibnamefont
  {Ortiz}},\ and\ \bibinfo {author} {\bibfnamefont {R.~L.}\ \bibnamefont
  {S{\'a}ez}},\ }\bibfield  {title} {\bibinfo {title} {Supersymmetry in the
  time domain and its applications in optics},\ }\href
  {https://doi.org/10.1038/s41467-020-14634-0} {\bibfield  {journal} {\bibinfo
  {journal} {Nature Communications}\ }\textbf {\bibinfo {volume} {11}},\
  \bibinfo {pages} {813} (\bibinfo {year} {2020})}\BibitemShut {NoStop}%
\bibitem [{\citenamefont {Yu}\ and\ \citenamefont
  {Yang}(2010)}]{yu2010simulating}%
  \BibitemOpen
  \bibfield  {author} {\bibinfo {author} {\bibfnamefont {Y.}~\bibnamefont
  {Yu}}\ and\ \bibinfo {author} {\bibfnamefont {K.}~\bibnamefont {Yang}},\
  }\bibfield  {title} {\bibinfo {title} {Simulating the {Wess}-{Zumino}
  {Supersymmetry} {Model} in {Optical} {Lattices}},\ }\href
  {https://doi.org/10.1103/PhysRevLett.105.150605} {\bibfield  {journal}
  {\bibinfo  {journal} {Physical Review Letters}\ }\textbf {\bibinfo {volume}
  {105}},\ \bibinfo {pages} {150605} (\bibinfo {year} {2010})}\BibitemShut
  {NoStop}%
\bibitem [{\citenamefont {Hokmabadi}\ \emph {et~al.}(2019)\citenamefont
  {Hokmabadi}, \citenamefont {Nye}, \citenamefont {El-Ganainy}, \citenamefont
  {Christodoulides},\ and\ \citenamefont
  {Khajavikhan}}]{hokmabadi2019supersymmetric}%
  \BibitemOpen
  \bibfield  {author} {\bibinfo {author} {\bibfnamefont {M.~P.}\ \bibnamefont
  {Hokmabadi}}, \bibinfo {author} {\bibfnamefont {N.~S.}\ \bibnamefont {Nye}},
  \bibinfo {author} {\bibfnamefont {R.}~\bibnamefont {El-Ganainy}}, \bibinfo
  {author} {\bibfnamefont {D.~N.}\ \bibnamefont {Christodoulides}},\ and\
  \bibinfo {author} {\bibfnamefont {M.}~\bibnamefont {Khajavikhan}},\
  }\bibfield  {title} {\bibinfo {title} {Supersymmetric laser arrays},\ }\href
  {https://doi.org/10.1126/science.aav5103} {\bibfield  {journal} {\bibinfo
  {journal} {Science}\ }\textbf {\bibinfo {volume} {363}},\ \bibinfo {pages}
  {623} (\bibinfo {year} {2019})}\BibitemShut {NoStop}%
\bibitem [{\citenamefont {Ma}\ \emph {et~al.}(2023)\citenamefont {Ma},
  \citenamefont {Jia}, \citenamefont {Bi}, \citenamefont {Ning}, \citenamefont
  {Guan}, \citenamefont {Liu}, \citenamefont {Wang},\ and\ \citenamefont
  {Zhang}}]{ma2023gaugefield}%
  \BibitemOpen
  \bibfield  {author} {\bibinfo {author} {\bibfnamefont {S.}~\bibnamefont
  {Ma}}, \bibinfo {author} {\bibfnamefont {H.}~\bibnamefont {Jia}}, \bibinfo
  {author} {\bibfnamefont {Y.}~\bibnamefont {Bi}}, \bibinfo {author}
  {\bibfnamefont {S.}~\bibnamefont {Ning}}, \bibinfo {author} {\bibfnamefont
  {F.}~\bibnamefont {Guan}}, \bibinfo {author} {\bibfnamefont {H.}~\bibnamefont
  {Liu}}, \bibinfo {author} {\bibfnamefont {C.}~\bibnamefont {Wang}},\ and\
  \bibinfo {author} {\bibfnamefont {S.}~\bibnamefont {Zhang}},\ }\bibfield
  {title} {\bibinfo {title} {Gauge {Field} {Induced} {Chiral} {Zero} {Mode} in
  {Five}-{Dimensional} {Yang} {Monopole} {Metamaterials}},\ }\href
  {https://doi.org/10.1103/PhysRevLett.130.243801} {\bibfield  {journal}
  {\bibinfo  {journal} {Physical Review Letters}\ }\textbf {\bibinfo {volume}
  {130}},\ \bibinfo {pages} {243801} (\bibinfo {year} {2023})}\BibitemShut
  {NoStop}%
\bibitem [{\citenamefont {Yu}\ and\ \citenamefont
  {Yang}(2008)}]{yu2008supersymmetry}%
  \BibitemOpen
  \bibfield  {author} {\bibinfo {author} {\bibfnamefont {Y.}~\bibnamefont
  {Yu}}\ and\ \bibinfo {author} {\bibfnamefont {K.}~\bibnamefont {Yang}},\
  }\bibfield  {title} {\bibinfo {title} {Supersymmetry and the
  {Goldstino}-{Like} {Mode} in {Bose}-{Fermi} {Mixtures}},\ }\href
  {https://doi.org/10.1103/PhysRevLett.100.090404} {\bibfield  {journal}
  {\bibinfo  {journal} {Physical Review Letters}\ }\textbf {\bibinfo {volume}
  {100}},\ \bibinfo {pages} {090404} (\bibinfo {year} {2008})}\BibitemShut
  {NoStop}%
\bibitem [{\citenamefont {Cai}\ \emph {et~al.}(2022)\citenamefont {Cai},
  \citenamefont {Wu}, \citenamefont {Mei}, \citenamefont {Zhao}, \citenamefont
  {Jiang}, \citenamefont {Yao}, \citenamefont {He}, \citenamefont {Zhou},\ and\
  \citenamefont {Duan}}]{cai2022observation}%
  \BibitemOpen
  \bibfield  {author} {\bibinfo {author} {\bibfnamefont {M.-L.}\ \bibnamefont
  {Cai}}, \bibinfo {author} {\bibfnamefont {Y.-K.}\ \bibnamefont {Wu}},
  \bibinfo {author} {\bibfnamefont {Q.-X.}\ \bibnamefont {Mei}}, \bibinfo
  {author} {\bibfnamefont {W.-D.}\ \bibnamefont {Zhao}}, \bibinfo {author}
  {\bibfnamefont {Y.}~\bibnamefont {Jiang}}, \bibinfo {author} {\bibfnamefont
  {L.}~\bibnamefont {Yao}}, \bibinfo {author} {\bibfnamefont {L.}~\bibnamefont
  {He}}, \bibinfo {author} {\bibfnamefont {Z.-C.}\ \bibnamefont {Zhou}},\ and\
  \bibinfo {author} {\bibfnamefont {L.-M.}\ \bibnamefont {Duan}},\ }\bibfield
  {title} {\bibinfo {title} {Observation of supersymmetry and its spontaneous
  breaking in a trapped ion quantum simulator},\ }\href
  {https://doi.org/10.1038/s41467-022-31058-0} {\bibfield  {journal} {\bibinfo
  {journal} {Nature Communications}\ }\textbf {\bibinfo {volume} {13}},\
  \bibinfo {pages} {3412} (\bibinfo {year} {2022})}\BibitemShut {NoStop}%
\bibitem [{\citenamefont {Min{\'a}{\v r}}\ \emph {et~al.}(2022)\citenamefont
  {Min{\'a}{\v r}}, \citenamefont {van Voorden},\ and\ \citenamefont
  {Schoutens}}]{minavr2022kinkdynamics}%
  \BibitemOpen
  \bibfield  {author} {\bibinfo {author} {\bibfnamefont {J.}~\bibnamefont
  {Min{\'a}{\v r}}}, \bibinfo {author} {\bibfnamefont {B.}~\bibnamefont {van
  Voorden}},\ and\ \bibinfo {author} {\bibfnamefont {K.}~\bibnamefont
  {Schoutens}},\ }\bibfield  {title} {\bibinfo {title} {Kink {Dynamics} and
  {Quantum} {Simulation} of {Supersymmetric} {Lattice} {Hamiltonians}},\ }\href
  {https://doi.org/10.1103/PhysRevLett.128.050504} {\bibfield  {journal}
  {\bibinfo  {journal} {Physical Review Letters}\ }\textbf {\bibinfo {volume}
  {128}},\ \bibinfo {pages} {050504} (\bibinfo {year} {2022})}\BibitemShut
  {NoStop}%
\bibitem [{\citenamefont {Grover}\ \emph {et~al.}(2014)\citenamefont {Grover},
  \citenamefont {Sheng},\ and\ \citenamefont
  {Vishwanath}}]{grover2014emergent}%
  \BibitemOpen
  \bibfield  {author} {\bibinfo {author} {\bibfnamefont {T.}~\bibnamefont
  {Grover}}, \bibinfo {author} {\bibfnamefont {D.~N.}\ \bibnamefont {Sheng}},\
  and\ \bibinfo {author} {\bibfnamefont {A.}~\bibnamefont {Vishwanath}},\
  }\bibfield  {title} {\bibinfo {title} {Emergent {Space}-{Time}
  {Supersymmetry} at the {Boundary} of a {Topological} {Phase}},\ }\href
  {https://doi.org/10.1126/science.1248253} {\bibfield  {journal} {\bibinfo
  {journal} {Science}\ }\textbf {\bibinfo {volume} {344}},\ \bibinfo {pages}
  {280} (\bibinfo {year} {2014})}\BibitemShut {NoStop}%
\bibitem [{\citenamefont {Jian}\ \emph {et~al.}(2015)\citenamefont {Jian},
  \citenamefont {Jiang},\ and\ \citenamefont {Yao}}]{jian2015emergent}%
  \BibitemOpen
  \bibfield  {author} {\bibinfo {author} {\bibfnamefont {S.-K.}\ \bibnamefont
  {Jian}}, \bibinfo {author} {\bibfnamefont {Y.-F.}\ \bibnamefont {Jiang}},\
  and\ \bibinfo {author} {\bibfnamefont {H.}~\bibnamefont {Yao}},\ }\bibfield
  {title} {\bibinfo {title} {Emergent {Spacetime} {Supersymmetry} in {3D}
  {Weyl} {Semimetals} and {2D} {Dirac} {Semimetals}},\ }\href
  {https://doi.org/10.1103/PhysRevLett.114.237001} {\bibfield  {journal}
  {\bibinfo  {journal} {Physical Review Letters}\ }\textbf {\bibinfo {volume}
  {114}},\ \bibinfo {pages} {237001} (\bibinfo {year} {2015})}\BibitemShut
  {NoStop}%
\bibitem [{\citenamefont {Li}\ \emph {et~al.}(2017)\citenamefont {Li},
  \citenamefont {Jiang},\ and\ \citenamefont {Yao}}]{li2017edgequantum}%
  \BibitemOpen
  \bibfield  {author} {\bibinfo {author} {\bibfnamefont {Z.-X.}\ \bibnamefont
  {Li}}, \bibinfo {author} {\bibfnamefont {Y.-F.}\ \bibnamefont {Jiang}},\ and\
  \bibinfo {author} {\bibfnamefont {H.}~\bibnamefont {Yao}},\ }\bibfield
  {title} {\bibinfo {title} {Edge {Quantum} {Criticality} and {Emergent}
  {Supersymmetry} in {Topological} {Phases}},\ }\href
  {https://doi.org/10.1103/PhysRevLett.119.107202} {\bibfield  {journal}
  {\bibinfo  {journal} {Physical Review Letters}\ }\textbf {\bibinfo {volume}
  {119}},\ \bibinfo {pages} {107202} (\bibinfo {year} {2017})}\BibitemShut
  {NoStop}%
\bibitem [{\citenamefont {Weber}\ \emph {et~al.}(2023)\citenamefont {Weber},
  \citenamefont {Pletyukhov}, \citenamefont {Hou}, \citenamefont {Kennes},
  \citenamefont {Klinovaja}, \citenamefont {Loss},\ and\ \citenamefont
  {Schoeller}}]{weber2023secondorder}%
  \BibitemOpen
  \bibfield  {author} {\bibinfo {author} {\bibfnamefont {C.~S.}\ \bibnamefont
  {Weber}}, \bibinfo {author} {\bibfnamefont {M.}~\bibnamefont {Pletyukhov}},
  \bibinfo {author} {\bibfnamefont {Z.}~\bibnamefont {Hou}}, \bibinfo {author}
  {\bibfnamefont {D.~M.}\ \bibnamefont {Kennes}}, \bibinfo {author}
  {\bibfnamefont {J.}~\bibnamefont {Klinovaja}}, \bibinfo {author}
  {\bibfnamefont {D.}~\bibnamefont {Loss}},\ and\ \bibinfo {author}
  {\bibfnamefont {H.}~\bibnamefont {Schoeller}},\ }\bibfield  {title} {\bibinfo
  {title} {Second-order topology and supersymmetry in two-dimensional
  topological insulators},\ }\href
  {https://doi.org/10.1103/PhysRevB.107.235402} {\bibfield  {journal} {\bibinfo
   {journal} {Physical Review B}\ }\textbf {\bibinfo {volume} {107}},\ \bibinfo
  {pages} {235402} (\bibinfo {year} {2023})}\BibitemShut {NoStop}%
\bibitem [{\citenamefont {Ponte}\ and\ \citenamefont
  {Lee}(2014)}]{ponte2014emergence}%
  \BibitemOpen
  \bibfield  {author} {\bibinfo {author} {\bibfnamefont {P.}~\bibnamefont
  {Ponte}}\ and\ \bibinfo {author} {\bibfnamefont {S.-S.}\ \bibnamefont
  {Lee}},\ }\bibfield  {title} {\bibinfo {title} {Emergence of supersymmetry on
  the surface of three-dimensional topological insulators},\ }\href
  {https://doi.org/10.1088/1367-2630/16/1/013044} {\bibfield  {journal}
  {\bibinfo  {journal} {New Journal of Physics}\ }\textbf {\bibinfo {volume}
  {16}},\ \bibinfo {pages} {013044} (\bibinfo {year} {2014})}\BibitemShut
  {NoStop}%
\bibitem [{\citenamefont {Yu}\ \emph {et~al.}(2022)\citenamefont {Yu},
  \citenamefont {Zhao}, \citenamefont {Jian},\ and\ \citenamefont
  {Pan}}]{yu2022emergent}%
  \BibitemOpen
  \bibfield  {author} {\bibinfo {author} {\bibfnamefont {X.-J.}\ \bibnamefont
  {Yu}}, \bibinfo {author} {\bibfnamefont {P.-L.}\ \bibnamefont {Zhao}},
  \bibinfo {author} {\bibfnamefont {S.-K.}\ \bibnamefont {Jian}},\ and\
  \bibinfo {author} {\bibfnamefont {Z.}~\bibnamefont {Pan}},\ }\bibfield
  {title} {\bibinfo {title} {Emergent space-time supersymmetry at disordered
  quantum critical points},\ }\href
  {https://doi.org/10.1103/PhysRevB.105.205140} {\bibfield  {journal} {\bibinfo
   {journal} {Physical Review B}\ }\textbf {\bibinfo {volume} {105}},\ \bibinfo
  {pages} {205140} (\bibinfo {year} {2022})}\BibitemShut {NoStop}%
\bibitem [{\citenamefont {Ma}\ \emph {et~al.}(2021)\citenamefont {Ma},
  \citenamefont {Wang},\ and\ \citenamefont {Yang}}]{ma2021realization}%
  \BibitemOpen
  \bibfield  {author} {\bibinfo {author} {\bibfnamefont {K.~K.~W.}\
  \bibnamefont {Ma}}, \bibinfo {author} {\bibfnamefont {R.}~\bibnamefont
  {Wang}},\ and\ \bibinfo {author} {\bibfnamefont {K.}~\bibnamefont {Yang}},\
  }\bibfield  {title} {\bibinfo {title} {Realization of {Supersymmetry} and
  {Its} {Spontaneous} {Breaking} in {Quantum} {Hall} {Edges}},\ }\href
  {https://doi.org/10.1103/PhysRevLett.126.206801} {\bibfield  {journal}
  {\bibinfo  {journal} {Physical Review Letters}\ }\textbf {\bibinfo {volume}
  {126}},\ \bibinfo {pages} {206801} (\bibinfo {year} {2021})}\BibitemShut
  {NoStop}%
\bibitem [{\citenamefont {Zhai}\ \emph {et~al.}(2024)\citenamefont {Zhai},
  \citenamefont {Lin},\ and\ \citenamefont {Yao}}]{zhaiSupersymmetry2024}%
  \BibitemOpen
  \bibfield  {author} {\bibinfo {author} {\bibfnamefont {D.}~\bibnamefont
  {Zhai}}, \bibinfo {author} {\bibfnamefont {Z.}~\bibnamefont {Lin}},\ and\
  \bibinfo {author} {\bibfnamefont {W.}~\bibnamefont {Yao}},\ }\bibfield
  {title} {\bibinfo {title} {Supersymmetry dictated topology in periodic gauge
  fields and realization in strained and twisted {{2D}} materials},\ }\href
  {https://doi.org/10.1088/1361-6633/ad77f0} {\bibfield  {journal} {\bibinfo
  {journal} {Reports on Progress in Physics}\ }\textbf {\bibinfo {volume}
  {87}},\ \bibinfo {pages} {108004} (\bibinfo {year} {2024})}\BibitemShut
  {NoStop}%
\bibitem [{\citenamefont {Ezawa}\ \emph {et~al.}(2023)\citenamefont {Ezawa},
  \citenamefont {Ishida}, \citenamefont {Ota},\ and\ \citenamefont
  {Iwamoto}}]{ezawa2023supersymmetric}%
  \BibitemOpen
  \bibfield  {author} {\bibinfo {author} {\bibfnamefont {M.}~\bibnamefont
  {Ezawa}}, \bibinfo {author} {\bibfnamefont {N.}~\bibnamefont {Ishida}},
  \bibinfo {author} {\bibfnamefont {Y.}~\bibnamefont {Ota}},\ and\ \bibinfo
  {author} {\bibfnamefont {S.}~\bibnamefont {Iwamoto}},\ }\bibfield  {title}
  {\bibinfo {title} {Supersymmetric non-{Hermitian} topological interface
  laser},\ }\href {https://doi.org/10.1103/PhysRevB.107.085302} {\bibfield
  {journal} {\bibinfo  {journal} {Physical Review B}\ }\textbf {\bibinfo
  {volume} {107}},\ \bibinfo {pages} {085302} (\bibinfo {year}
  {2023})}\BibitemShut {NoStop}%
\bibitem [{\citenamefont {Prakash}\ and\ \citenamefont
  {Wang}(2021)}]{prakash2021boundary}%
  \BibitemOpen
  \bibfield  {author} {\bibinfo {author} {\bibfnamefont {A.}~\bibnamefont
  {Prakash}}\ and\ \bibinfo {author} {\bibfnamefont {J.}~\bibnamefont {Wang}},\
  }\bibfield  {title} {\bibinfo {title} {Boundary {Supersymmetry} of (1+1)d
  {Fermionic} {Symmetry}-{Protected} {Topological} {Phases}},\ }\href
  {https://doi.org/10.1103/PhysRevLett.126.236802} {\bibfield  {journal}
  {\bibinfo  {journal} {Physical Review Letters}\ }\textbf {\bibinfo {volume}
  {126}},\ \bibinfo {pages} {236802} (\bibinfo {year} {2021})}\BibitemShut
  {NoStop}%
\bibitem [{\citenamefont {Turzillo}\ and\ \citenamefont
  {You}(2021)}]{turzillo2021supersymmetric}%
  \BibitemOpen
  \bibfield  {author} {\bibinfo {author} {\bibfnamefont {A.}~\bibnamefont
  {Turzillo}}\ and\ \bibinfo {author} {\bibfnamefont {M.}~\bibnamefont {You}},\
  }\bibfield  {title} {\bibinfo {title} {Supersymmetric {Boundaries} of
  {One}-{Dimensional} {Phases} of {Fermions} beyond {Symmetry}-{Protected}
  {Topological} {States}},\ }\href
  {https://doi.org/10.1103/PhysRevLett.127.026402} {\bibfield  {journal}
  {\bibinfo  {journal} {Physical Review Letters}\ }\textbf {\bibinfo {volume}
  {127}},\ \bibinfo {pages} {026402} (\bibinfo {year} {2021})}\BibitemShut
  {NoStop}%
\bibitem [{\citenamefont {Behrends}\ and\ \citenamefont
  {B{\'e}ri}(2020)}]{behrends2020supersymmetry}%
  \BibitemOpen
  \bibfield  {author} {\bibinfo {author} {\bibfnamefont {J.}~\bibnamefont
  {Behrends}}\ and\ \bibinfo {author} {\bibfnamefont {B.}~\bibnamefont
  {B{\'e}ri}},\ }\bibfield  {title} {\bibinfo {title} {Supersymmetry in the
  {Standard} {Sachdev}-{Ye}-{Kitaev} {Model}},\ }\href
  {https://doi.org/10.1103/PhysRevLett.124.236804} {\bibfield  {journal}
  {\bibinfo  {journal} {Physical Review Letters}\ }\textbf {\bibinfo {volume}
  {124}},\ \bibinfo {pages} {236804} (\bibinfo {year} {2020})}\BibitemShut
  {NoStop}%
\bibitem [{\citenamefont {Nayga}\ \emph {et~al.}(2019)\citenamefont {Nayga},
  \citenamefont {Rachel},\ and\ \citenamefont {Vojta}}]{nayga2019magnonlandau}%
  \BibitemOpen
  \bibfield  {author} {\bibinfo {author} {\bibfnamefont {M.~M.}\ \bibnamefont
  {Nayga}}, \bibinfo {author} {\bibfnamefont {S.}~\bibnamefont {Rachel}},\ and\
  \bibinfo {author} {\bibfnamefont {M.}~\bibnamefont {Vojta}},\ }\bibfield
  {title} {\bibinfo {title} {Magnon {Landau} {Levels} and {Emergent}
  {Supersymmetry} in {Strained} {Antiferromagnets}},\ }\href
  {https://doi.org/10.1103/PhysRevLett.123.207204} {\bibfield  {journal}
  {\bibinfo  {journal} {Physical Review Letters}\ }\textbf {\bibinfo {volume}
  {123}},\ \bibinfo {pages} {207204} (\bibinfo {year} {2019})}\BibitemShut
  {NoStop}%
\bibitem [{\citenamefont {Zhang}\ \emph {et~al.}(2005)\citenamefont {Zhang},
  \citenamefont {Tan}, \citenamefont {Stormer},\ and\ \citenamefont
  {Kim}}]{zhang-graphene-2005}%
  \BibitemOpen
  \bibfield  {author} {\bibinfo {author} {\bibfnamefont {Y.}~\bibnamefont
  {Zhang}}, \bibinfo {author} {\bibfnamefont {Y.-W.}\ \bibnamefont {Tan}},
  \bibinfo {author} {\bibfnamefont {H.~L.}\ \bibnamefont {Stormer}},\ and\
  \bibinfo {author} {\bibfnamefont {P.}~\bibnamefont {Kim}},\ }\bibfield
  {title} {\bibinfo {title} {Experimental observation of the quantum {{Hall}}
  effect and {{Berry}}'s phase in graphene},\ }\href
  {https://doi.org/10.1038/nature04235} {\bibfield  {journal} {\bibinfo
  {journal} {Nature}\ }\textbf {\bibinfo {volume} {438}},\ \bibinfo {pages}
  {201} (\bibinfo {year} {2005})}\BibitemShut {NoStop}%
\bibitem [{\citenamefont {Li}\ \emph {et~al.}(2016{\natexlab{a}})\citenamefont
  {Li}, \citenamefont {Kharzeev}, \citenamefont {Zhang}, \citenamefont {Huang},
  \citenamefont {Pletikosi{\'c}}, \citenamefont {Fedorov}, \citenamefont
  {Zhong}, \citenamefont {Schneeloch}, \citenamefont {Gu},\ and\ \citenamefont
  {Valla}}]{li-weyl-2016}%
  \BibitemOpen
  \bibfield  {author} {\bibinfo {author} {\bibfnamefont {Q.}~\bibnamefont
  {Li}}, \bibinfo {author} {\bibfnamefont {D.~E.}\ \bibnamefont {Kharzeev}},
  \bibinfo {author} {\bibfnamefont {C.}~\bibnamefont {Zhang}}, \bibinfo
  {author} {\bibfnamefont {Y.}~\bibnamefont {Huang}}, \bibinfo {author}
  {\bibfnamefont {I.}~\bibnamefont {Pletikosi{\'c}}}, \bibinfo {author}
  {\bibfnamefont {A.~V.}\ \bibnamefont {Fedorov}}, \bibinfo {author}
  {\bibfnamefont {R.~D.}\ \bibnamefont {Zhong}}, \bibinfo {author}
  {\bibfnamefont {J.~A.}\ \bibnamefont {Schneeloch}}, \bibinfo {author}
  {\bibfnamefont {G.~D.}\ \bibnamefont {Gu}},\ and\ \bibinfo {author}
  {\bibfnamefont {T.}~\bibnamefont {Valla}},\ }\bibfield  {title} {\bibinfo
  {title} {Chiral magnetic effect in {{ZrTe5}}},\ }\href
  {https://doi.org/10.1038/nphys3648} {\bibfield  {journal} {\bibinfo
  {journal} {Nature Physics}\ }\textbf {\bibinfo {volume} {12}},\ \bibinfo
  {pages} {550} (\bibinfo {year} {2016}{\natexlab{a}})}\BibitemShut {NoStop}%
\bibitem [{\citenamefont {Huang}\ \emph {et~al.}(2015)\citenamefont {Huang},
  \citenamefont {Zhao}, \citenamefont {Long}, \citenamefont {Wang},
  \citenamefont {Chen}, \citenamefont {Yang}, \citenamefont {Liang},
  \citenamefont {Xue}, \citenamefont {Weng}, \citenamefont {Fang},
  \citenamefont {Dai},\ and\ \citenamefont {Chen}}]{Huang-prx-2015}%
  \BibitemOpen
  \bibfield  {author} {\bibinfo {author} {\bibfnamefont {X.}~\bibnamefont
  {Huang}}, \bibinfo {author} {\bibfnamefont {L.}~\bibnamefont {Zhao}},
  \bibinfo {author} {\bibfnamefont {Y.}~\bibnamefont {Long}}, \bibinfo {author}
  {\bibfnamefont {P.}~\bibnamefont {Wang}}, \bibinfo {author} {\bibfnamefont
  {D.}~\bibnamefont {Chen}}, \bibinfo {author} {\bibfnamefont {Z.}~\bibnamefont
  {Yang}}, \bibinfo {author} {\bibfnamefont {H.}~\bibnamefont {Liang}},
  \bibinfo {author} {\bibfnamefont {M.}~\bibnamefont {Xue}}, \bibinfo {author}
  {\bibfnamefont {H.}~\bibnamefont {Weng}}, \bibinfo {author} {\bibfnamefont
  {Z.}~\bibnamefont {Fang}}, \bibinfo {author} {\bibfnamefont {X.}~\bibnamefont
  {Dai}},\ and\ \bibinfo {author} {\bibfnamefont {G.}~\bibnamefont {Chen}},\
  }\bibfield  {title} {\bibinfo {title} {Observation of the
  chiral-anomaly-induced negative magnetoresistance in 3d weyl semimetal
  taas},\ }\href {https://doi.org/10.1103/PhysRevX.5.031023} {\bibfield
  {journal} {\bibinfo  {journal} {Phys. Rev. X}\ }\textbf {\bibinfo {volume}
  {5}},\ \bibinfo {pages} {031023} (\bibinfo {year} {2015})}\BibitemShut
  {NoStop}%
\bibitem [{\citenamefont {Zhao}\ \emph {et~al.}(2015)\citenamefont {Zhao},
  \citenamefont {Liu}, \citenamefont {Zhang}, \citenamefont {Wang},
  \citenamefont {Wang}, \citenamefont {Lin}, \citenamefont {Xing},
  \citenamefont {Lu}, \citenamefont {Liu}, \citenamefont {Wang}, \citenamefont
  {Brombosz}, \citenamefont {Xiao}, \citenamefont {Jia}, \citenamefont {Xie},\
  and\ \citenamefont {Wang}}]{Zhao-prx-2015}%
  \BibitemOpen
  \bibfield  {author} {\bibinfo {author} {\bibfnamefont {Y.}~\bibnamefont
  {Zhao}}, \bibinfo {author} {\bibfnamefont {H.}~\bibnamefont {Liu}}, \bibinfo
  {author} {\bibfnamefont {C.}~\bibnamefont {Zhang}}, \bibinfo {author}
  {\bibfnamefont {H.}~\bibnamefont {Wang}}, \bibinfo {author} {\bibfnamefont
  {J.}~\bibnamefont {Wang}}, \bibinfo {author} {\bibfnamefont {Z.}~\bibnamefont
  {Lin}}, \bibinfo {author} {\bibfnamefont {Y.}~\bibnamefont {Xing}}, \bibinfo
  {author} {\bibfnamefont {H.}~\bibnamefont {Lu}}, \bibinfo {author}
  {\bibfnamefont {J.}~\bibnamefont {Liu}}, \bibinfo {author} {\bibfnamefont
  {Y.}~\bibnamefont {Wang}}, \bibinfo {author} {\bibfnamefont {S.~M.}\
  \bibnamefont {Brombosz}}, \bibinfo {author} {\bibfnamefont {Z.}~\bibnamefont
  {Xiao}}, \bibinfo {author} {\bibfnamefont {S.}~\bibnamefont {Jia}}, \bibinfo
  {author} {\bibfnamefont {X.~C.}\ \bibnamefont {Xie}},\ and\ \bibinfo {author}
  {\bibfnamefont {J.}~\bibnamefont {Wang}},\ }\bibfield  {title} {\bibinfo
  {title} {Anisotropic fermi surface and quantum limit transport in high
  mobility three-dimensional dirac semimetal
  ${\mathrm{cd}}_{3}{\mathrm{as}}_{2}$},\ }\href
  {https://doi.org/10.1103/PhysRevX.5.031037} {\bibfield  {journal} {\bibinfo
  {journal} {Phys. Rev. X}\ }\textbf {\bibinfo {volume} {5}},\ \bibinfo {pages}
  {031037} (\bibinfo {year} {2015})}\BibitemShut {NoStop}%
\bibitem [{\citenamefont {Zhao}\ and\ \citenamefont {Wang}(2015)}]{zhao-weyl}%
  \BibitemOpen
  \bibfield  {author} {\bibinfo {author} {\bibfnamefont {Y.~X.}\ \bibnamefont
  {Zhao}}\ and\ \bibinfo {author} {\bibfnamefont {Z.~D.}\ \bibnamefont
  {Wang}},\ }\bibfield  {title} {\bibinfo {title} {Disordered weyl semimetals
  and their topological family},\ }\href
  {https://doi.org/10.1103/PhysRevLett.114.206602} {\bibfield  {journal}
  {\bibinfo  {journal} {Phys. Rev. Lett.}\ }\textbf {\bibinfo {volume} {114}},\
  \bibinfo {pages} {206602} (\bibinfo {year} {2015})}\BibitemShut {NoStop}%
\bibitem [{\citenamefont {Shekhar}\ \emph {et~al.}(2015)\citenamefont
  {Shekhar}, \citenamefont {Nayak}, \citenamefont {Sun}, \citenamefont
  {Schmidt}, \citenamefont {Nicklas}, \citenamefont {Leermakers}, \citenamefont
  {Zeitler}, \citenamefont {Skourski}, \citenamefont {Wosnitza}, \citenamefont
  {Liu}, \citenamefont {Chen}, \citenamefont {Schnelle}, \citenamefont
  {Borrmann}, \citenamefont {Grin}, \citenamefont {Felser},\ and\ \citenamefont
  {Yan}}]{shekhar-weyl-2015}%
  \BibitemOpen
  \bibfield  {author} {\bibinfo {author} {\bibfnamefont {C.}~\bibnamefont
  {Shekhar}}, \bibinfo {author} {\bibfnamefont {A.~K.}\ \bibnamefont {Nayak}},
  \bibinfo {author} {\bibfnamefont {Y.}~\bibnamefont {Sun}}, \bibinfo {author}
  {\bibfnamefont {M.}~\bibnamefont {Schmidt}}, \bibinfo {author} {\bibfnamefont
  {M.}~\bibnamefont {Nicklas}}, \bibinfo {author} {\bibfnamefont
  {I.}~\bibnamefont {Leermakers}}, \bibinfo {author} {\bibfnamefont
  {U.}~\bibnamefont {Zeitler}}, \bibinfo {author} {\bibfnamefont
  {Y.}~\bibnamefont {Skourski}}, \bibinfo {author} {\bibfnamefont
  {J.}~\bibnamefont {Wosnitza}}, \bibinfo {author} {\bibfnamefont
  {Z.}~\bibnamefont {Liu}}, \bibinfo {author} {\bibfnamefont {Y.}~\bibnamefont
  {Chen}}, \bibinfo {author} {\bibfnamefont {W.}~\bibnamefont {Schnelle}},
  \bibinfo {author} {\bibfnamefont {H.}~\bibnamefont {Borrmann}}, \bibinfo
  {author} {\bibfnamefont {Y.}~\bibnamefont {Grin}}, \bibinfo {author}
  {\bibfnamefont {C.}~\bibnamefont {Felser}},\ and\ \bibinfo {author}
  {\bibfnamefont {B.}~\bibnamefont {Yan}},\ }\bibfield  {title} {\bibinfo
  {title} {Extremely large magnetoresistance and ultrahigh mobility in the
  topological {{Weyl}} semimetal candidate {{NbP}}},\ }\href
  {https://doi.org/10.1038/nphys3372} {\bibfield  {journal} {\bibinfo
  {journal} {Nature Physics}\ }\textbf {\bibinfo {volume} {11}},\ \bibinfo
  {pages} {645} (\bibinfo {year} {2015})}\BibitemShut {NoStop}%
\bibitem [{\citenamefont {Wang}\ \emph {et~al.}(2017)\citenamefont {Wang},
  \citenamefont {Sun}, \citenamefont {Lu},\ and\ \citenamefont
  {Xie}}]{wang3DQuantum2017}%
  \BibitemOpen
  \bibfield  {author} {\bibinfo {author} {\bibfnamefont {C.~M.}\ \bibnamefont
  {Wang}}, \bibinfo {author} {\bibfnamefont {H.-P.}\ \bibnamefont {Sun}},
  \bibinfo {author} {\bibfnamefont {H.-Z.}\ \bibnamefont {Lu}},\ and\ \bibinfo
  {author} {\bibfnamefont {X.~C.}\ \bibnamefont {Xie}},\ }\bibfield  {title}
  {\bibinfo {title} {{{3D Quantum Hall Effect}} of {{Fermi Arcs}} in
  {{Topological Semimetals}}},\ }\href
  {https://doi.org/10.1103/PhysRevLett.119.136806} {\bibfield  {journal}
  {\bibinfo  {journal} {Physical Review Letters}\ }\textbf {\bibinfo {volume}
  {119}},\ \bibinfo {pages} {136806} (\bibinfo {year} {2017})}\BibitemShut
  {NoStop}%
\bibitem [{\citenamefont {Rui}\ \emph {et~al.}(2022)\citenamefont {Rui},
  \citenamefont {Zheng}, \citenamefont {Wang},\ and\ \citenamefont
  {Wang}}]{rui-weyl}%
  \BibitemOpen
  \bibfield  {author} {\bibinfo {author} {\bibfnamefont {W.~B.}\ \bibnamefont
  {Rui}}, \bibinfo {author} {\bibfnamefont {Z.}~\bibnamefont {Zheng}}, \bibinfo
  {author} {\bibfnamefont {C.}~\bibnamefont {Wang}},\ and\ \bibinfo {author}
  {\bibfnamefont {Z.~D.}\ \bibnamefont {Wang}},\ }\bibfield  {title} {\bibinfo
  {title} {Non-hermitian spatial symmetries and their stabilized normal and
  exceptional topological semimetals},\ }\href
  {https://doi.org/10.1103/PhysRevLett.128.226401} {\bibfield  {journal}
  {\bibinfo  {journal} {Phys. Rev. Lett.}\ }\textbf {\bibinfo {volume} {128}},\
  \bibinfo {pages} {226401} (\bibinfo {year} {2022})}\BibitemShut {NoStop}%
\bibitem [{\citenamefont {Yuan}\ \emph {et~al.}(2018)\citenamefont {Yuan},
  \citenamefont {Yan}, \citenamefont {Song}, \citenamefont {Zhang},
  \citenamefont {Li}, \citenamefont {Zhang}, \citenamefont {Liu}, \citenamefont
  {Wang}, \citenamefont {Zhao}, \citenamefont {Lin}, \citenamefont {Xie},
  \citenamefont {Ludwig}, \citenamefont {Jiang}, \citenamefont {Zhang},
  \citenamefont {Shang}, \citenamefont {Ye}, \citenamefont {Wang},
  \citenamefont {Chen}, \citenamefont {Xia}, \citenamefont {Smirnov},
  \citenamefont {Chen}, \citenamefont {Wang}, \citenamefont {Yan},\ and\
  \citenamefont {Xiu}}]{yuanChirals2018a}%
  \BibitemOpen
  \bibfield  {author} {\bibinfo {author} {\bibfnamefont {X.}~\bibnamefont
  {Yuan}}, \bibinfo {author} {\bibfnamefont {Z.}~\bibnamefont {Yan}}, \bibinfo
  {author} {\bibfnamefont {C.}~\bibnamefont {Song}}, \bibinfo {author}
  {\bibfnamefont {M.}~\bibnamefont {Zhang}}, \bibinfo {author} {\bibfnamefont
  {Z.}~\bibnamefont {Li}}, \bibinfo {author} {\bibfnamefont {C.}~\bibnamefont
  {Zhang}}, \bibinfo {author} {\bibfnamefont {Y.}~\bibnamefont {Liu}}, \bibinfo
  {author} {\bibfnamefont {W.}~\bibnamefont {Wang}}, \bibinfo {author}
  {\bibfnamefont {M.}~\bibnamefont {Zhao}}, \bibinfo {author} {\bibfnamefont
  {Z.}~\bibnamefont {Lin}}, \bibinfo {author} {\bibfnamefont {T.}~\bibnamefont
  {Xie}}, \bibinfo {author} {\bibfnamefont {J.}~\bibnamefont {Ludwig}},
  \bibinfo {author} {\bibfnamefont {Y.}~\bibnamefont {Jiang}}, \bibinfo
  {author} {\bibfnamefont {X.}~\bibnamefont {Zhang}}, \bibinfo {author}
  {\bibfnamefont {C.}~\bibnamefont {Shang}}, \bibinfo {author} {\bibfnamefont
  {Z.}~\bibnamefont {Ye}}, \bibinfo {author} {\bibfnamefont {J.}~\bibnamefont
  {Wang}}, \bibinfo {author} {\bibfnamefont {F.}~\bibnamefont {Chen}}, \bibinfo
  {author} {\bibfnamefont {Z.}~\bibnamefont {Xia}}, \bibinfo {author}
  {\bibfnamefont {D.}~\bibnamefont {Smirnov}}, \bibinfo {author} {\bibfnamefont
  {X.}~\bibnamefont {Chen}}, \bibinfo {author} {\bibfnamefont {Z.}~\bibnamefont
  {Wang}}, \bibinfo {author} {\bibfnamefont {H.}~\bibnamefont {Yan}},\ and\
  \bibinfo {author} {\bibfnamefont {F.}~\bibnamefont {Xiu}},\ }\bibfield
  {title} {\bibinfo {title} {Chiral {{Landau}} levels in {{Weyl}} semimetal
  {{NbAs}} with multiple topological carriers},\ }\href
  {https://doi.org/10.1038/s41467-018-04080-4} {\bibfield  {journal} {\bibinfo
  {journal} {Nature Communications}\ }\textbf {\bibinfo {volume} {9}},\
  \bibinfo {pages} {1854} (\bibinfo {year} {2018})}\BibitemShut {NoStop}%
\bibitem [{\citenamefont {Kim}\ \emph {et~al.}(2018)\citenamefont {Kim},
  \citenamefont {Seo}, \citenamefont {Lee}, \citenamefont {Ko}, \citenamefont
  {Kim}, \citenamefont {Jang}, \citenamefont {Ok}, \citenamefont {Lee},
  \citenamefont {Jo}, \citenamefont {Kang}, \citenamefont {Shim}, \citenamefont
  {Kim}, \citenamefont {Yeom}, \citenamefont {Il~Min}, \citenamefont {Yang},\
  and\ \citenamefont {Kim}}]{kim-nodal-line-2018}%
  \BibitemOpen
  \bibfield  {author} {\bibinfo {author} {\bibfnamefont {K.}~\bibnamefont
  {Kim}}, \bibinfo {author} {\bibfnamefont {J.}~\bibnamefont {Seo}}, \bibinfo
  {author} {\bibfnamefont {E.}~\bibnamefont {Lee}}, \bibinfo {author}
  {\bibfnamefont {K.-T.}\ \bibnamefont {Ko}}, \bibinfo {author} {\bibfnamefont
  {B.~S.}\ \bibnamefont {Kim}}, \bibinfo {author} {\bibfnamefont {B.~G.}\
  \bibnamefont {Jang}}, \bibinfo {author} {\bibfnamefont {J.~M.}\ \bibnamefont
  {Ok}}, \bibinfo {author} {\bibfnamefont {J.}~\bibnamefont {Lee}}, \bibinfo
  {author} {\bibfnamefont {Y.~J.}\ \bibnamefont {Jo}}, \bibinfo {author}
  {\bibfnamefont {W.}~\bibnamefont {Kang}}, \bibinfo {author} {\bibfnamefont
  {J.~H.}\ \bibnamefont {Shim}}, \bibinfo {author} {\bibfnamefont
  {C.}~\bibnamefont {Kim}}, \bibinfo {author} {\bibfnamefont {H.~W.}\
  \bibnamefont {Yeom}}, \bibinfo {author} {\bibfnamefont {B.}~\bibnamefont
  {Il~Min}}, \bibinfo {author} {\bibfnamefont {B.-J.}\ \bibnamefont {Yang}},\
  and\ \bibinfo {author} {\bibfnamefont {J.~S.}\ \bibnamefont {Kim}},\
  }\bibfield  {title} {\bibinfo {title} {Large anomalous {{Hall}} current
  induced by topological nodal lines in a ferromagnetic van der {{Waals}}
  semimetal},\ }\href {https://doi.org/10.1038/s41563-018-0132-3} {\bibfield
  {journal} {\bibinfo  {journal} {Nature Materials}\ }\textbf {\bibinfo
  {volume} {17}},\ \bibinfo {pages} {794} (\bibinfo {year} {2018})}\BibitemShut
  {NoStop}%
\bibitem [{\citenamefont {Zhang}\ \emph {et~al.}(2017)\citenamefont {Zhang},
  \citenamefont {Xu}, \citenamefont {Wang}, \citenamefont {Lin}, \citenamefont
  {Du}, \citenamefont {Guo}, \citenamefont {Lee}, \citenamefont {Lu},
  \citenamefont {Feng}, \citenamefont {Huang}, \citenamefont {Chang},
  \citenamefont {Hsu}, \citenamefont {Liu}, \citenamefont {Lin}, \citenamefont
  {Li}, \citenamefont {Zhang}, \citenamefont {Zhang}, \citenamefont {Xie},
  \citenamefont {Neupert}, \citenamefont {Hasan}, \citenamefont {Lu},
  \citenamefont {Wang},\ and\ \citenamefont
  {Jia}}]{zhang2017magnetictunnellinginduced}%
  \BibitemOpen
  \bibfield  {author} {\bibinfo {author} {\bibfnamefont {C.-L.}\ \bibnamefont
  {Zhang}}, \bibinfo {author} {\bibfnamefont {S.-Y.}\ \bibnamefont {Xu}},
  \bibinfo {author} {\bibfnamefont {C.~M.}\ \bibnamefont {Wang}}, \bibinfo
  {author} {\bibfnamefont {Z.}~\bibnamefont {Lin}}, \bibinfo {author}
  {\bibfnamefont {Z.~Z.}\ \bibnamefont {Du}}, \bibinfo {author} {\bibfnamefont
  {C.}~\bibnamefont {Guo}}, \bibinfo {author} {\bibfnamefont {C.-C.}\
  \bibnamefont {Lee}}, \bibinfo {author} {\bibfnamefont {H.}~\bibnamefont
  {Lu}}, \bibinfo {author} {\bibfnamefont {Y.}~\bibnamefont {Feng}}, \bibinfo
  {author} {\bibfnamefont {S.-M.}\ \bibnamefont {Huang}}, \bibinfo {author}
  {\bibfnamefont {G.}~\bibnamefont {Chang}}, \bibinfo {author} {\bibfnamefont
  {C.-H.}\ \bibnamefont {Hsu}}, \bibinfo {author} {\bibfnamefont
  {H.}~\bibnamefont {Liu}}, \bibinfo {author} {\bibfnamefont {H.}~\bibnamefont
  {Lin}}, \bibinfo {author} {\bibfnamefont {L.}~\bibnamefont {Li}}, \bibinfo
  {author} {\bibfnamefont {C.}~\bibnamefont {Zhang}}, \bibinfo {author}
  {\bibfnamefont {J.}~\bibnamefont {Zhang}}, \bibinfo {author} {\bibfnamefont
  {X.-C.}\ \bibnamefont {Xie}}, \bibinfo {author} {\bibfnamefont
  {T.}~\bibnamefont {Neupert}}, \bibinfo {author} {\bibfnamefont {M.~Z.}\
  \bibnamefont {Hasan}}, \bibinfo {author} {\bibfnamefont {H.-Z.}\ \bibnamefont
  {Lu}}, \bibinfo {author} {\bibfnamefont {J.}~\bibnamefont {Wang}},\ and\
  \bibinfo {author} {\bibfnamefont {S.}~\bibnamefont {Jia}},\ }\bibfield
  {title} {\bibinfo {title} {Magnetic-tunnelling-induced {Weyl} node
  annihilation in {TaP}},\ }\href {https://doi.org/10.1038/nphys4183}
  {\bibfield  {journal} {\bibinfo  {journal} {Nature Physics}\ }\textbf
  {\bibinfo {volume} {13}},\ \bibinfo {pages} {979} (\bibinfo {year}
  {2017})}\BibitemShut {NoStop}%
\bibitem [{\citenamefont {Ramshaw}\ \emph {et~al.}(2018)\citenamefont
  {Ramshaw}, \citenamefont {Modic}, \citenamefont {Shekhter}, \citenamefont
  {Zhang}, \citenamefont {Kim}, \citenamefont {Moll}, \citenamefont {Bachmann},
  \citenamefont {Chan}, \citenamefont {Betts}, \citenamefont {Balakirev},
  \citenamefont {Migliori}, \citenamefont {Ghimire}, \citenamefont {Bauer},
  \citenamefont {Ronning},\ and\ \citenamefont
  {McDonald}}]{ramshaw2018quantum}%
  \BibitemOpen
  \bibfield  {author} {\bibinfo {author} {\bibfnamefont {B.~J.}\ \bibnamefont
  {Ramshaw}}, \bibinfo {author} {\bibfnamefont {K.~A.}\ \bibnamefont {Modic}},
  \bibinfo {author} {\bibfnamefont {A.}~\bibnamefont {Shekhter}}, \bibinfo
  {author} {\bibfnamefont {Y.}~\bibnamefont {Zhang}}, \bibinfo {author}
  {\bibfnamefont {E.-A.}\ \bibnamefont {Kim}}, \bibinfo {author} {\bibfnamefont
  {P.~J.~W.}\ \bibnamefont {Moll}}, \bibinfo {author} {\bibfnamefont {M.~D.}\
  \bibnamefont {Bachmann}}, \bibinfo {author} {\bibfnamefont {M.~K.}\
  \bibnamefont {Chan}}, \bibinfo {author} {\bibfnamefont {J.~B.}\ \bibnamefont
  {Betts}}, \bibinfo {author} {\bibfnamefont {F.}~\bibnamefont {Balakirev}},
  \bibinfo {author} {\bibfnamefont {A.}~\bibnamefont {Migliori}}, \bibinfo
  {author} {\bibfnamefont {N.~J.}\ \bibnamefont {Ghimire}}, \bibinfo {author}
  {\bibfnamefont {E.~D.}\ \bibnamefont {Bauer}}, \bibinfo {author}
  {\bibfnamefont {F.}~\bibnamefont {Ronning}},\ and\ \bibinfo {author}
  {\bibfnamefont {R.~D.}\ \bibnamefont {McDonald}},\ }\bibfield  {title}
  {\bibinfo {title} {Quantum limit transport and destruction of the {Weyl}
  nodes in {TaAs}},\ }\href {https://doi.org/10.1038/s41467-018-04542-9}
  {\bibfield  {journal} {\bibinfo  {journal} {Nature Communications}\ }\textbf
  {\bibinfo {volume} {9}},\ \bibinfo {pages} {2217} (\bibinfo {year}
  {2018})}\BibitemShut {NoStop}%
\bibitem [{Foo()}]{Footnote-q}%
  \BibitemOpen
  \href@noop {} {}\bibinfo {note} {Our analysis is restricted to square
  matrices for $q$, while more general rectangular matrices are considered in
  Ref.~\citep{Roychowdhury-SUSY}.}\BibitemShut {Stop}%
\bibitem [{Sup()}]{SuppInf}%
  \BibitemOpen
  \href@noop {} {\bibinfo  {journal} {{See the Supplemental Material for
  details}}\ }\BibitemShut {NoStop}%
\bibitem [{\citenamefont {Castro~Neto}\ \emph {et~al.}(2009)\citenamefont
  {Castro~Neto}, \citenamefont {Guinea}, \citenamefont {Peres}, \citenamefont
  {Novoselov},\ and\ \citenamefont {Geim}}]{castroneto-Graphene-2009}%
  \BibitemOpen
\bibfield  {journal} {  }\bibfield  {author} {\bibinfo {author} {\bibfnamefont
  {A.~H.}\ \bibnamefont {Castro~Neto}}, \bibinfo {author} {\bibfnamefont
  {F.}~\bibnamefont {Guinea}}, \bibinfo {author} {\bibfnamefont {N.~M.~R.}\
  \bibnamefont {Peres}}, \bibinfo {author} {\bibfnamefont {K.~S.}\ \bibnamefont
  {Novoselov}},\ and\ \bibinfo {author} {\bibfnamefont {A.~K.}\ \bibnamefont
  {Geim}},\ }\bibfield  {title} {\bibinfo {title} {The electronic properties of
  graphene},\ }\href {https://doi.org/10.1103/RevModPhys.81.109} {\bibfield
  {journal} {\bibinfo  {journal} {Reviews of Modern Physics}\ }\textbf
  {\bibinfo {volume} {81}},\ \bibinfo {pages} {109} (\bibinfo {year}
  {2009})}\BibitemShut {NoStop}%
\bibitem [{\citenamefont {Mari{\~n}o}(2015)}]{marino2015instantons}%
  \BibitemOpen
  \bibfield  {author} {\bibinfo {author} {\bibfnamefont {M.}~\bibnamefont
  {Mari{\~n}o}},\ }\href {https://doi.org/10.1017/CBO9781107705968} {\emph
  {\bibinfo {title} {Instantons and {Large} {N}: {An} {Introduction} to
  {Non}-{Perturbative} {Methods} in {Quantum} {Field} {Theory}}}}\ (\bibinfo
  {publisher} {Cambridge University Press},\ \bibinfo {address} {Cambridge},\
  \bibinfo {year} {2015})\BibitemShut {NoStop}%
\bibitem [{\citenamefont {van Holten}(1983)}]{vanholten1983instantons}%
  \BibitemOpen
  \bibfield  {author} {\bibinfo {author} {\bibfnamefont {J.~W.}\ \bibnamefont
  {van Holten}},\ }\bibfield  {title} {\bibinfo {title} {Instantons in
  {Supersymmetric} {Quantum} {Mechanics}},\ }in\ \href
  {https://doi.org/10.1007/978-1-4615-9299-0_32} {\emph {\bibinfo {booktitle}
  {Unification of {Fundamental} {Particle} {Interactions} {II}}}},\ \bibinfo
  {series and number} {Ettore {Majorana} {International} {Science} {Series}},\
  \bibinfo {editor} {edited by\ \bibinfo {editor} {\bibfnamefont
  {J.}~\bibnamefont {Ellis}}\ and\ \bibinfo {editor} {\bibfnamefont
  {S.}~\bibnamefont {Ferrara}}}\ (\bibinfo  {publisher} {Springer US},\
  \bibinfo {address} {Boston, MA},\ \bibinfo {year} {1983})\ pp.\ \bibinfo
  {pages} {505--513}\BibitemShut {NoStop}%
\bibitem [{foo()}]{footnote}%
  \BibitemOpen
  \href@noop {} {}\bibinfo {note} {When the perpendicular magnetic field is
  applied in the $z$-direction, using the same replacement as in
  Eq.~\eqref{eq:magnetic-field}, the Hamiltonian becomes
  $H_{\text{w}}=-iv_{x}\partial_{\mathsf{x}}\sigma_{1}+w_{\text{double}}(\mathsf{x})\sigma_{2}+v_{z}k_{z}\sigma_{3}$.
  The spectrum must be gapped when $k_{z}\neq0$, since the $k_{z}$ term
  anticommutes with others.}\BibitemShut {Stop}%
\bibitem [{Foo()}]{Footnote3}%
  \BibitemOpen
  \href@noop {} {}\bibinfo {note} {The reported "threshold" is when the inverse
  of magnetic length,i.e., $l_B^{-1}$, is comparable to the separation $\Delta
  K=2K$. It can also be explained by our instanton calculation. Express the
  energy gap as $\Delta \propto \exp[-\mu_y/(6v_x)\Delta K^{3} l_B^2]$. The
  region for the gap to be not exponentially small gap occurs when the exponent
  is comparable to $-1$. It leads to $l_B^{-1}=\sqrt{\mu_y/(6v_x)}\Delta
  K^{3/2}$, which roughly agrees with the observation $l_B^{-1}\sim \Delta
  K$.}\BibitemShut {Stop}%
\bibitem [{\citenamefont {Li}\ and\ \citenamefont
  {Andrei}(2007)}]{liLandauLevels2007}%
  \BibitemOpen
  \bibfield  {author} {\bibinfo {author} {\bibfnamefont {G.}~\bibnamefont
  {Li}}\ and\ \bibinfo {author} {\bibfnamefont {E.~Y.}\ \bibnamefont
  {Andrei}},\ }\bibfield  {title} {\bibinfo {title} {Observation of {{Landau}}
  levels of {{Dirac}} fermions in graphite},\ }\href
  {https://doi.org/10.1038/nphys653} {\bibfield  {journal} {\bibinfo  {journal}
  {Nature Physics}\ }\textbf {\bibinfo {volume} {3}},\ \bibinfo {pages} {623}
  (\bibinfo {year} {2007})}\BibitemShut {NoStop}%
\bibitem [{\citenamefont {Jeon}\ \emph {et~al.}(2014)\citenamefont {Jeon},
  \citenamefont {Zhou}, \citenamefont {Gyenis}, \citenamefont {Feldman},
  \citenamefont {Kimchi}, \citenamefont {Potter}, \citenamefont {Gibson},
  \citenamefont {Cava}, \citenamefont {Vishwanath},\ and\ \citenamefont
  {Yazdani}}]{jeon2014}%
  \BibitemOpen
  \bibfield  {author} {\bibinfo {author} {\bibfnamefont {S.}~\bibnamefont
  {Jeon}}, \bibinfo {author} {\bibfnamefont {B.~B.}\ \bibnamefont {Zhou}},
  \bibinfo {author} {\bibfnamefont {A.}~\bibnamefont {Gyenis}}, \bibinfo
  {author} {\bibfnamefont {B.~E.}\ \bibnamefont {Feldman}}, \bibinfo {author}
  {\bibfnamefont {I.}~\bibnamefont {Kimchi}}, \bibinfo {author} {\bibfnamefont
  {A.~C.}\ \bibnamefont {Potter}}, \bibinfo {author} {\bibfnamefont {Q.~D.}\
  \bibnamefont {Gibson}}, \bibinfo {author} {\bibfnamefont {R.~J.}\
  \bibnamefont {Cava}}, \bibinfo {author} {\bibfnamefont {A.}~\bibnamefont
  {Vishwanath}},\ and\ \bibinfo {author} {\bibfnamefont {A.}~\bibnamefont
  {Yazdani}},\ }\bibfield  {title} {\bibinfo {title} {Landau quantization and
  quasiparticle interference in the three-dimensional {{Dirac}} semimetal
  {{Cd3As2}}},\ }\href {https://doi.org/10.1038/nmat4023} {\bibfield  {journal}
  {\bibinfo  {journal} {Nature Materials}\ }\textbf {\bibinfo {volume} {13}},\
  \bibinfo {pages} {851} (\bibinfo {year} {2014})}\BibitemShut {NoStop}%
\bibitem [{\citenamefont {Schwenk}\ \emph {et~al.}(2020)\citenamefont
  {Schwenk}, \citenamefont {Kim}, \citenamefont {Berwanger}, \citenamefont
  {Ghahari}, \citenamefont {Walkup}, \citenamefont {Slot}, \citenamefont {Le},
  \citenamefont {Cullen}, \citenamefont {Blankenship}, \citenamefont
  {Vranjkovic}, \citenamefont {Hug}, \citenamefont {Kuk}, \citenamefont
  {Giessibl},\ and\ \citenamefont {Stroscio}}]{schwenk-STM-2020}%
  \BibitemOpen
  \bibfield  {author} {\bibinfo {author} {\bibfnamefont {J.}~\bibnamefont
  {Schwenk}}, \bibinfo {author} {\bibfnamefont {S.}~\bibnamefont {Kim}},
  \bibinfo {author} {\bibfnamefont {J.}~\bibnamefont {Berwanger}}, \bibinfo
  {author} {\bibfnamefont {F.}~\bibnamefont {Ghahari}}, \bibinfo {author}
  {\bibfnamefont {D.}~\bibnamefont {Walkup}}, \bibinfo {author} {\bibfnamefont
  {M.~R.}\ \bibnamefont {Slot}}, \bibinfo {author} {\bibfnamefont {S.~T.}\
  \bibnamefont {Le}}, \bibinfo {author} {\bibfnamefont {W.~G.}\ \bibnamefont
  {Cullen}}, \bibinfo {author} {\bibfnamefont {S.~R.}\ \bibnamefont
  {Blankenship}}, \bibinfo {author} {\bibfnamefont {S.}~\bibnamefont
  {Vranjkovic}}, \bibinfo {author} {\bibfnamefont {H.~J.}\ \bibnamefont {Hug}},
  \bibinfo {author} {\bibfnamefont {Y.}~\bibnamefont {Kuk}}, \bibinfo {author}
  {\bibfnamefont {F.~J.}\ \bibnamefont {Giessibl}},\ and\ \bibinfo {author}
  {\bibfnamefont {J.~A.}\ \bibnamefont {Stroscio}},\ }\bibfield  {title}
  {\bibinfo {title} {Achieving {{$\mu$eV}} tunneling resolution in an
  in-operando scanning tunneling microscopy, atomic force microscopy, and
  magnetotransport system for quantum materials research},\ }\href
  {https://doi.org/10.1063/5.0005320} {\bibfield  {journal} {\bibinfo
  {journal} {Review of Scientific Instruments}\ }\textbf {\bibinfo {volume}
  {91}},\ \bibinfo {pages} {071101} (\bibinfo {year} {2020})}\BibitemShut
  {NoStop}%
\bibitem [{\citenamefont {Fang}\ \emph {et~al.}(2015)\citenamefont {Fang},
  \citenamefont {Chen}, \citenamefont {Kee},\ and\ \citenamefont
  {Fu}}]{chen-nl-2015}%
  \BibitemOpen
  \bibfield  {author} {\bibinfo {author} {\bibfnamefont {C.}~\bibnamefont
  {Fang}}, \bibinfo {author} {\bibfnamefont {Y.}~\bibnamefont {Chen}}, \bibinfo
  {author} {\bibfnamefont {H.-Y.}\ \bibnamefont {Kee}},\ and\ \bibinfo {author}
  {\bibfnamefont {L.}~\bibnamefont {Fu}},\ }\bibfield  {title} {\bibinfo
  {title} {Topological nodal line semimetals with and without spin-orbital
  coupling},\ }\href {https://doi.org/10.1103/PhysRevB.92.081201} {\bibfield
  {journal} {\bibinfo  {journal} {Phys. Rev. B}\ }\textbf {\bibinfo {volume}
  {92}},\ \bibinfo {pages} {081201} (\bibinfo {year} {2015})}\BibitemShut
  {NoStop}%
\bibitem [{\citenamefont {Li}\ \emph {et~al.}(2016{\natexlab{b}})\citenamefont
  {Li}, \citenamefont {Ma}, \citenamefont {Cheng}, \citenamefont {Wang},
  \citenamefont {Li}, \citenamefont {Zhang}, \citenamefont {Li},\ and\
  \citenamefont {Chen}}]{Li-nodal-line}%
  \BibitemOpen
  \bibfield  {author} {\bibinfo {author} {\bibfnamefont {R.}~\bibnamefont
  {Li}}, \bibinfo {author} {\bibfnamefont {H.}~\bibnamefont {Ma}}, \bibinfo
  {author} {\bibfnamefont {X.}~\bibnamefont {Cheng}}, \bibinfo {author}
  {\bibfnamefont {S.}~\bibnamefont {Wang}}, \bibinfo {author} {\bibfnamefont
  {D.}~\bibnamefont {Li}}, \bibinfo {author} {\bibfnamefont {Z.}~\bibnamefont
  {Zhang}}, \bibinfo {author} {\bibfnamefont {Y.}~\bibnamefont {Li}},\ and\
  \bibinfo {author} {\bibfnamefont {X.-Q.}\ \bibnamefont {Chen}},\ }\bibfield
  {title} {\bibinfo {title} {Dirac node lines in pure alkali earth metals},\
  }\href {https://doi.org/10.1103/PhysRevLett.117.096401} {\bibfield  {journal}
  {\bibinfo  {journal} {Phys. Rev. Lett.}\ }\textbf {\bibinfo {volume} {117}},\
  \bibinfo {pages} {096401} (\bibinfo {year} {2016}{\natexlab{b}})}\BibitemShut
  {NoStop}%
\bibitem [{\citenamefont {Rui}\ \emph {et~al.}(2018)\citenamefont {Rui},
  \citenamefont {Zhao},\ and\ \citenamefont {Schnyder}}]{rui2018topological}%
  \BibitemOpen
  \bibfield  {author} {\bibinfo {author} {\bibfnamefont {W.~B.}\ \bibnamefont
  {Rui}}, \bibinfo {author} {\bibfnamefont {Y.~X.}\ \bibnamefont {Zhao}},\ and\
  \bibinfo {author} {\bibfnamefont {A.~P.}\ \bibnamefont {Schnyder}},\
  }\bibfield  {title} {\bibinfo {title} {Topological transport in {Dirac}
  nodal-line semimetals},\ }\href {https://doi.org/10.1103/PhysRevB.97.161113}
  {\bibfield  {journal} {\bibinfo  {journal} {Physical Review B}\ }\textbf
  {\bibinfo {volume} {97}},\ \bibinfo {pages} {161113} (\bibinfo {year}
  {2018})}\BibitemShut {NoStop}%
\bibitem [{\citenamefont {Chiu}\ \emph {et~al.}(2016)\citenamefont {Chiu},
  \citenamefont {Teo}, \citenamefont {Schnyder},\ and\ \citenamefont
  {Ryu}}]{Chiu_RMP_2016}%
  \BibitemOpen
  \bibfield  {author} {\bibinfo {author} {\bibfnamefont {C.-K.}\ \bibnamefont
  {Chiu}}, \bibinfo {author} {\bibfnamefont {J.~C.~Y.}\ \bibnamefont {Teo}},
  \bibinfo {author} {\bibfnamefont {A.~P.}\ \bibnamefont {Schnyder}},\ and\
  \bibinfo {author} {\bibfnamefont {S.}~\bibnamefont {Ryu}},\ }\bibfield
  {title} {\bibinfo {title} {Classification of topological quantum matter with
  symmetries},\ }\href {https://doi.org/10.1103/RevModPhys.88.035005}
  {\bibfield  {journal} {\bibinfo  {journal} {Rev. Mod. Phys.}\ }\textbf
  {\bibinfo {volume} {88}},\ \bibinfo {pages} {035005} (\bibinfo {year}
  {2016})}\BibitemShut {NoStop}%
\bibitem [{\citenamefont {Tarnopolsky}\ \emph {et~al.}(2019)\citenamefont
  {Tarnopolsky}, \citenamefont {Kruchkov},\ and\ \citenamefont
  {Vishwanath}}]{chiral-limit-tbg}%
  \BibitemOpen
  \bibfield  {author} {\bibinfo {author} {\bibfnamefont {G.}~\bibnamefont
  {Tarnopolsky}}, \bibinfo {author} {\bibfnamefont {A.~J.}\ \bibnamefont
  {Kruchkov}},\ and\ \bibinfo {author} {\bibfnamefont {A.}~\bibnamefont
  {Vishwanath}},\ }\bibfield  {title} {\bibinfo {title} {Origin of magic angles
  in twisted bilayer graphene},\ }\href
  {https://doi.org/10.1103/PhysRevLett.122.106405} {\bibfield  {journal}
  {\bibinfo  {journal} {Phys. Rev. Lett.}\ }\textbf {\bibinfo {volume} {122}},\
  \bibinfo {pages} {106405} (\bibinfo {year} {2019})}\BibitemShut {NoStop}%
\bibitem [{\citenamefont {Wang}\ \emph {et~al.}(2021)\citenamefont {Wang},
  \citenamefont {Zheng}, \citenamefont {Millis},\ and\ \citenamefont
  {Cano}}]{Chiral-approximation-tbg}%
  \BibitemOpen
  \bibfield  {author} {\bibinfo {author} {\bibfnamefont {J.}~\bibnamefont
  {Wang}}, \bibinfo {author} {\bibfnamefont {Y.}~\bibnamefont {Zheng}},
  \bibinfo {author} {\bibfnamefont {A.~J.}\ \bibnamefont {Millis}},\ and\
  \bibinfo {author} {\bibfnamefont {J.}~\bibnamefont {Cano}},\ }\bibfield
  {title} {\bibinfo {title} {Chiral approximation to twisted bilayer graphene:
  Exact intravalley inversion symmetry, nodal structure, and implications for
  higher magic angles},\ }\href
  {https://doi.org/10.1103/PhysRevResearch.3.023155} {\bibfield  {journal}
  {\bibinfo  {journal} {Phys. Rev. Res.}\ }\textbf {\bibinfo {volume} {3}},\
  \bibinfo {pages} {023155} (\bibinfo {year} {2021})}\BibitemShut {NoStop}%
\bibitem [{\citenamefont {Fern{\'a}ndez~C.}\ and\ \citenamefont
  {Fern{\'a}ndez-Garc{\'i}a}(2004)}]{HigherorderSupersymmetricQuantum2004}%
  \BibitemOpen
  \bibfield  {author} {\bibinfo {author} {\bibfnamefont {D.~J.}\ \bibnamefont
  {Fern{\'a}ndez~C.}}\ and\ \bibinfo {author} {\bibfnamefont {N.}~\bibnamefont
  {Fern{\'a}ndez-Garc{\'i}a}},\ }\bibfield  {title} {\bibinfo {title}
  {Higher-order supersymmetric quantum mechanics},\ }\href
  {https://doi.org/10.1063/1.1853203} {\bibfield  {journal} {\bibinfo
  {journal} {AIP Conference Proceedings}\ }\textbf {\bibinfo {volume} {744}},\
  \bibinfo {pages} {236} (\bibinfo {year} {2004})}\BibitemShut {NoStop}%
\bibitem [{\citenamefont {Roychowdhury}\ \emph {et~al.}(2024)\citenamefont
  {Roychowdhury}, \citenamefont {Attig}, \citenamefont {Trebst},\ and\
  \citenamefont {Lawler}}]{Roychowdhury-SUSY}%
  \BibitemOpen
  \bibfield  {author} {\bibinfo {author} {\bibfnamefont {K.}~\bibnamefont
  {Roychowdhury}}, \bibinfo {author} {\bibfnamefont {J.}~\bibnamefont {Attig}},
  \bibinfo {author} {\bibfnamefont {S.}~\bibnamefont {Trebst}},\ and\ \bibinfo
  {author} {\bibfnamefont {M.~J.}\ \bibnamefont {Lawler}},\ }\bibfield  {title}
  {\bibinfo {title} {Supersymmetry on the lattice: Geometry, topology, and flat
  bands},\ }\href {https://doi.org/10.1103/PhysRevResearch.6.043273} {\bibfield
   {journal} {\bibinfo  {journal} {Phys. Rev. Res.}\ }\textbf {\bibinfo
  {volume} {6}},\ \bibinfo {pages} {043273} (\bibinfo {year}
  {2024})}\BibitemShut {NoStop}%
\end{thebibliography}%

\end{document}